%% file: final-tex-arxiv.tex
\documentclass[review,onefignum,onetabnum]{siamonline171218}

\usepackage{mathrsfs,bigints,mathtools,graphicx}

\input{ex_shared}

\ifpdf
\hypersetup{
	pdftitle={A two-dimensional swarmalator model with higher-order interactions},
	pdfauthor={Md Sayeed Anwar, Gourab Kumar Sar, Timoteo Carletti and Dibakar Ghosh}
}
\fi

\begin{document}
	
 \maketitle

\begin{abstract}
We study a simple two-dimensional swarmalator model that incorporates higher-order phase interactions, uncovering a diverse range of collective states. The latter include spatially coherent and gas-like configurations, neither of which appear in models with only pairwise interactions. Additionally, we discover bistability between various states, a phenomenon that arises directly from the inclusion of higher-order interactions. By analyzing several of these emergent states analytically, both for identical and nonidentical populations of swarmalators, we gain deeper insights into their underlying mechanisms and stability conditions. Our findings broaden the understanding of swarmalator dynamics and open new avenues for exploring complex collective behaviors in systems governed by higher-order interactions. 
\end{abstract}



\section{Introduction}
Swarmalators represent an intriguing class of interconnected moving entities (oscillators) that simultaneously synchronize their internal states and adjust their spatial positions in response to one another \cite{o2017oscillators}. This dual dynamical nature leads to a rich variety of emergent behaviors that bridge the gap between self-organized motion (swarming) \cite{darnton2010dynamics,fetecau2011swarm} and coordinated temporal dynamics (synchronization) \cite{pikovsky2003synchronization,strogatz2004sync}. Although swarmalators have been a relatively recent addition to the study of complex systems, they have proven to be useful for modeling systems that combine self-assembly and synchronization, ranging from biological systems to microrobots \cite{tamm1975role,verberck2022wavy,yan2012linking,o2019review,zhang2020reconfigurable,bricard2015emergent,riedel2005self,sar2024solvable}.

\par However, existing studies on swarmalators have largely focused on the understanding of their behavior through the lens of pairwise interactions among the entities composing the swarm \cite{sar2022dynamics,o2017oscillators,hong2023swarmalators,anwar2024forced,sar2023pinning,sar2023swarmalators,ghosh2023antiphase,ghosh2024amplitude,sar2022swarmalators}. Specifically, these studies assume that the overall interplay among the swarmalators is exhaustively described by combinations of spatial and dynamical interactions occurring in pairs. While this approach has offered useful insights, it overlooks an essential characteristic of many real-world systems: interactions often involve groups, not just pairs \cite{battiston2020networks,battiston2021physics,majhi2022dynamics}. For instance, in microbial populations, quorum sensing and collective decision-making often involve interactions that extend beyond pairs, where group-level coordination dictates the population's spatial and temporal dynamics \cite{mukherjee2019bacterial,zhou2020regulatory,ludington2022higher,swain2022higher,li2012quorum}. Beyond biology, higher-order interactions can play a critical role in artificial systems such as robotic swarms \cite{brutschy2014self,calderon2022swarm}. Here, coordination often requires multi-agent task allocation, where the behavior of one robot depends on the collective state of a group, rather than individual pairwise interactions.

\par Incorporating higher-order (group) interactions into the traditional swarmalator framework is, therefore, a natural and necessary step toward a more comprehensive understanding of these systems. Higher-order interactions profoundly reshape the landscape of collective behavior, by introducing novel dynamical regimes and altering stability properties. While such interactions have been explored in the contexts of synchronization and network dynamics \cite{anwar2022stability,anwar2024global,anwar2024synchronization,gambuzza2021stability,lucas2020multiorder}, their influence on swarmalator systems remains largely unexplored. A preliminary study has recently addressed this gap, by extending the framework as to include higher-order interactions among populations of swarmalators confined to move on a ring \cite{anwar2024collective}. While there are swarmalators, such as frogs, nematodes, and other organisms, that tend to favor the ring-like boundaries of their confined geometries \cite{aihara2014spatio,creppy2016symmetry,yuan2015hydrodynamic}, most physical systems involve swarmalators moving on higher-dimensional surfaces, such as in two or three dimensions. In this work, we thus aim to incorporate higher-order interactions in a relatively more general case where swarmalators are confined to move in two spatial dimensions. Indeed, physical systems, such as active colloids \cite{pohl2014dynamic} or sperm \cite{rothschild1963non}, are naturally attracted to move along two-dimensional surfaces.

\par The goal of this work is two-fold. First, we aim to explore how higher-order interactions influence the collective dynamics of swarmalators confined to two-dimensional surfaces. Second, we seek to develop a mathematically tractable framework that incorporates group-level interactions among the swarmalators. To this end, we consider a model in which swarmalators are subject to higher-order interactions (specifically simultaneous interactions up to groups of three) and move within a two-dimensional space with periodic boundary conditions, rather than an open plane. The use of periodic boundaries facilitates analytical treatment of the system's emergent dynamics, a common approach in active matter research and related fields \cite{vicsek1995novel,ginelli2016physics}.

\par With these simplifications, we are able to derive exact expressions for the stability and bifurcations of several collective states. Notably, our analysis reveals that the inclusion of higher-order interactions leads to the emergence of both stationary and nonstationary long-term states, which do not arise when only pairwise interactions are considered. Moreover, we find that higher-order interactions introduce bistability between distinct states, contributing to the persistence of the synchronization (sync) state even in the scenario where pairwise coupling is repulsive. Additionally, we observe that higher-order interactions induce abrupt transitions between emergent states, highlighting their significant role in shaping the overall system dynamics.

\section{The model}
In order to incorporate higher-order interactions within the populations of swarmalators moving on a two-dimensional surface, we consider the following model
\begin{subequations}\label{model}    
\begin{equation} \label{sw_position_xeq}
\begin{array}{cc}
    \dot{x}_{i}= u_{i}+\dfrac{J}{N} \sum\limits_{j=1}^{N} \sin(x_{j}-x_{i}) \cos(\theta_{j}-\theta_{i}),
\end{array}
\end{equation}
\begin{equation} \label{sw_position_yeq}
\begin{array}{cc}
    \dot{y}_{i}= v_{i}+\dfrac{J}{N} \sum\limits_{j=1}^{N} \sin(y_{j}-y_{i}) \cos(\theta_{j}-\theta_{i}),
\end{array}
\end{equation}
\begin{equation}  \label{sw_phase_eq}
\begin{array}{cc}
    \dot{\theta}_{i}= \omega_{i}+\dfrac{K_1}{N} \sum\limits_{j=1}^{N} \sin(\theta_{j}-\theta_{i}) \big[\cos(x_{j}-x_{i})+\cos(y_{j}-y_{i}) \big] \\ +\dfrac{K_2}{N^2} \sum\limits_{j=1}^{N} \sum\limits_{k=1}^{N} \sin(2\theta_{j}-\theta_{k}-\theta_{i}) \big[\cos(2x_{j}-x_{k}-x_{i}) +\cos(2y_{j}-y_{k}-y_{i})\big], 
\end{array}
\end{equation}
\end{subequations}
where $\mathbf{x}_{i}= (x_i,y_i)$ represents the position of the $i$-th swarmalator, subject to periodic boundary condition $(x_i,y_i) \in \mathbb{S}^{1} \times \mathbb{S}^{1}$ and $\theta_{i} \in \mathbb{S}^{1}$ represents its internal phase. Moreover $(u_i,v_i)$ are the free velocities and $\omega_{i}$ the natural frequencies of the uncoupled swarmalators. $J$ denotes the coupling strength associated with the space interaction, while $K_1$ and $K_2$ control the pairwise and higher-order phase interactions, respectively. The parameter $J$ regulates space-space coupling, where $J>0$ means the swarmalators are inclined to spatially synchronize, while $J<0$ implies spatial repulsion. $K_{1}$ and $K_{2}$ have analogous behavior but for the phase interactions. Notably, when $K_{2}=0$, the model reduces to the recently introduced analytically tractable pairwise two-dimensional swarmalator model \cite{o2024solvable}. As in the pairwise model, we do not account for spatial hard-shell repulsion, and the spatial dynamics reflect sync-dependent self-assembly characterized by the spatial attraction term $\sin(\mathbf{x}_{j}-\mathbf{x}_{i})$ and modulated by the phase differences $\cos(\theta_{j}-\theta_{i})$. On the other hand, the phase dynamics does the opposite and encodes position dependent synchrony. It combines both pairwise and triadic interactions, where the sine terms minimize phase differences between individual swarmalators as they interact pairwise or in a group of three, with distance kernels $\cos({x}_{j}-{x}_{i})+\cos({y}_{j}-{y}_{i}) $ and $\cos(2{x}_{j}-{x}_{k}-{x}_{i})+\cos(2{y}_{j}-{y}_{k}-{y}_{i})$. Here, we introduce higher-order interactions in the phase dynamics of
the swarmalators, aiming to exert a direct influence only
on their individual phases and not on their spatial positions. Nevertheless, the group interactions influence the spatial dynamics through its impact on the coupling term in Eq.~\eqref{sw_phase_eq}.

\par Let us observe that the chosen form of higher-order interactions, i.e., $g(2x_{j}-x_{k}-x_{i})$, is not unique. One could have chosen the alternative form, $g(x_{j}+x_{k}-2x_{i})$, as studied in the literature of Kuramoto oscillators with higher-order interactions \cite{skardal2020higher,anwar2024self}. However, such an alternative introduces analytical complexity, particularly when dealing with non-identical populations of swarmalators.

\par Before proceeding further, we rewrite the model \eqref{model} by converting the trigonometric expressions to complex exponential 
\begin{subequations}\label{model-op}    
	\begin{align} \label{xeq}
			\dot{x}_{i}= u_{i}+\frac{J}{2} \text{Im}\left[ W_1^+ e^{-i(x_i + \theta_i)} + W_1^- e^{-i(x_i - \theta_i)}\right],
	\end{align}
	\begin{align} \label{yeq}	
			\dot{y}_{i}= v_{i} + \frac{J}{2} \text{Im}\left[ Z_1^+ e^{-i(y_i + \theta_i)} + Z_1^- e^{-i(y_i - \theta_i)}\right],
	\end{align}
	\begin{align}\label{thetaeq}
		\dot{\theta}_{i} = & \omega_{i} + \frac{K_1}{2} \text{Im}\left[ W_1^+ e^{-i(x_i + \theta_i)} - W_1^- e^{-i(x_i - \theta_i)}\right] + \frac{K_1}{2} \text{Im}\left[ Z_1^+ e^{-i(y_i + \theta_i)} - Z_1^- e^{-i(y_i - \theta_i)}\right] \nonumber \\  &+ \frac{K_2}{2} \text{Im}\left[ W_2^+ (W_1^+)^* e^{-i(x_i + \theta_i)} - W_2^- (W_1^-)^* e^{-i(x_i - \theta_i)}\right] + \frac{K_2}{2} \text{Im}\left[ Z_2^+ (Z_1^+)^* e^{-i(y_i + \theta_i)} - Z_2^- (Z_1^-)^* e^{-i(y_i - \theta_i)}\right], 
	\end{align}
\end{subequations}
where $^*$ denotes complex conjugate, $\text{Im}[z]$ represents imaginary part of a complex number $z$ and the complex order parameters are given by
\begin{align}
	W_m^{\pm}  = \frac{1}{N} \sum_{j} e^{ i m(x_j \pm \theta_j)}, \\
	Z_m^{\pm}  = \frac{1}{N} \sum_{j} e^{ i m(y_j \pm \theta_j)},& \;\;m=1,2.
\end{align}
Equation \eqref{model-op} thus introduces a set of complex order parameters $W_m^{\pm}=S_m^{\pm} e^{i \phi_m^{\pm}}$ and $Z_m^{\pm}=T_m^{\pm} e^{i \psi_m^{\pm}}$ that measure the ``space-phase order'' of the whole population. Here, $S_m^{\pm}$ ($T_m^{\pm}$) represent the amplitudes, while $\phi_m^{\pm}$ ($\psi_m^{\pm}$) denote the arguments of the respective order parameters. We redefine $S_m^{+}$ as the maximum between $S_m^{+}$ and $S_m^{-}$, and similarly, $T_m^{+}$ as the maximum between $T_m^{+}$ and $T_m^{-}$. When the space and phase coordinates of the swarmalators are fully uncorrelated (i.e., uniformly distributed), the order parameters $S_{1}^{\pm} (T_{1}^{\pm})$ reach their minimum value of zero, in the limit of unbounded system size. Conversely, when the system is fully synchronized in both space and phase (i.e., $x_{i}=x^{*},y_{i}=y^{*},\; \text{and}\; \theta_{i}=\theta^{*}$ for all $i$), the order parameters reach their maximum value of one. Therefore, the order parameters quantify the degree of correlation between spatial and phase coordinates, ranging from $0$ (no correlation) to $1$ (perfect sync). The coupling dependencies of these order parameters in Eq.~\eqref{model-op} suggests that varying the coupling strengths may lead to different combinations of $(S_{1}^{+},S_{1}^{-},T_{1}^{+},T_{1}^{+})$ and thus we expect the system to display a rich array of dynamical behaviors. Additionally, due to the presence of higher-order interactions, analogous to higher-order Kuramoto model \cite{skardal2020higher}, we can define another set of order parameters $(S_{2}^{\pm},T_{2}^{\pm})$ that could capture clustering or other forms of group dynamics. In the present study, however, we will explore the system dynamics based on the order parameters $(S_{1}^{\pm},T_{1}^{\pm})$ only.

\section{Identical swarmalators}
To begin, we first consider the simplified scenario in which the free velocities and internal frequencies of all swarmalators are identical, i.e., $(u_i,v_i,\omega_i)= (u,v,\omega)$ for all $i$. Then by choosing an appropriate reference frame, we set $u=v=\omega=0$ without loss of generality. This assumption reduces the complexity of the system and allows us to focus on the effects of the coupling terms and higher-order interactions.  
\begin{figure*}[hpt] 
	\centerline
	{\includegraphics[scale=0.12]{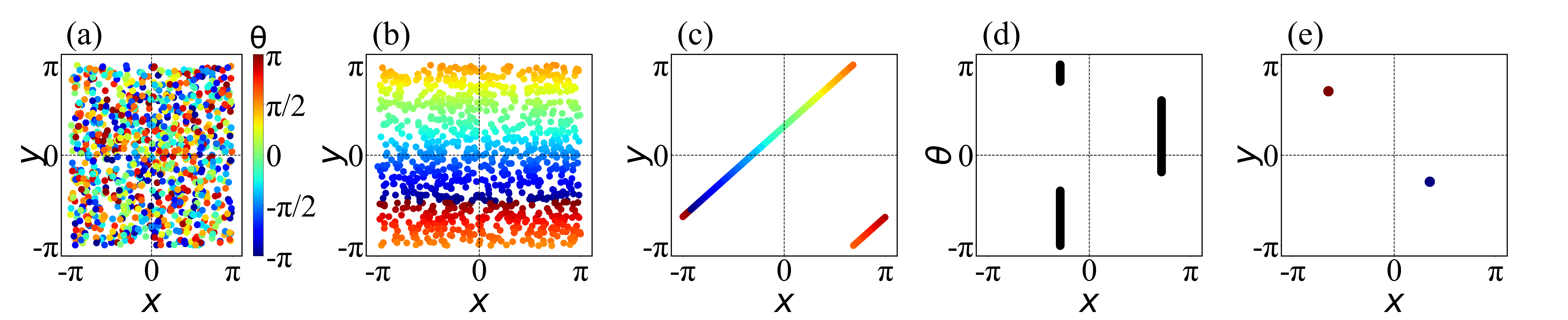}}
	\centerline
	{\includegraphics[scale=0.12]{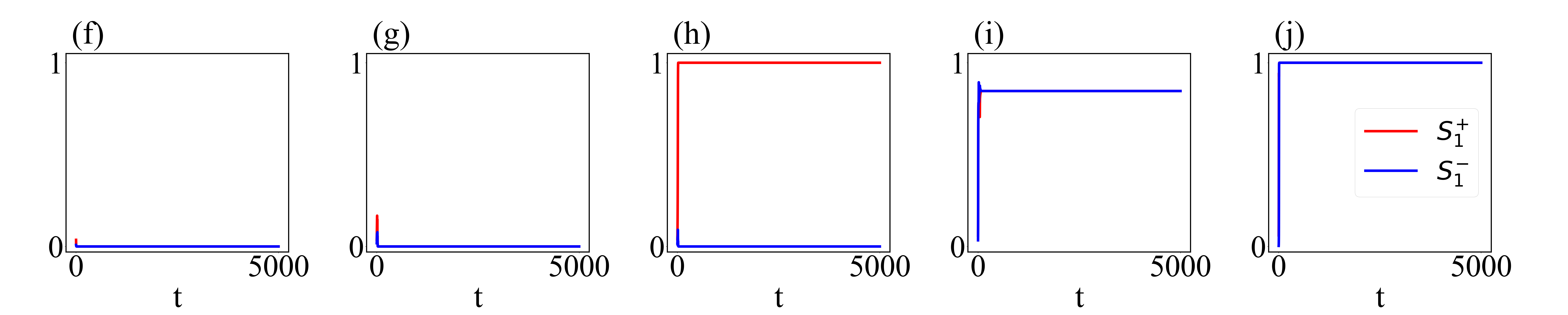}}
	\centerline
	{\includegraphics[scale=0.12]{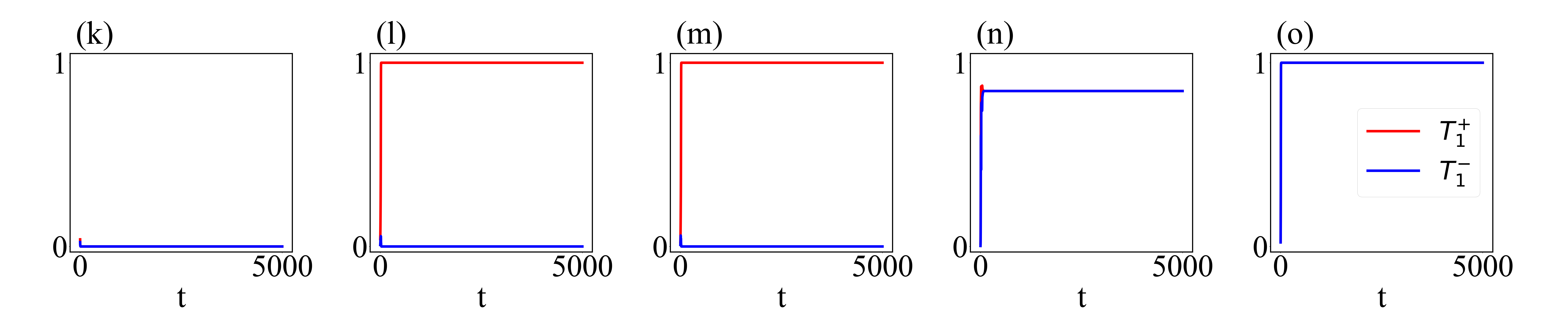}}
	\caption{{\bf Stationary states.} Top row: snapshots of the states in $(x,y)$ plane with colors indicating the phase $\theta$, except panel (d) where the snapshot of the spatially coherent state is portrayed in $(x,\theta)$ plane. Middle row: variation of order parameters $S_{1}^{\pm}$ as a function of time, and bottom row: order parameters $T_{1}^{\pm}$ as a function of time. (a, f, k) Async state: $(J, K_1, K_2)=(1, -3, 1)$, (b, g, l) Thick phase wave state: $(J, K_1, K_2)=(-1, 2, 1)$, (c, h, m) Thin phase wave state: $(J, K_1, K_2)=(2, -1, 1)$, (d, i, n) Spatially coherent state: $(J, K_1, K_2)=(1, 2, -4)$, and (e, j, o) Sync state: $(J, K_1, K_2)=(1, 2, 1)$. States are obtained by integrating Eq.~\eqref{model} with $N=1000$ swarmalators for a total of $T=5000$ time units using an adaptive Julia Ode solver having relative tolerance of $10^{-8}$.}
	\label{static_states}
\end{figure*}
\subsection{Numerical results}
We performed dedicated numerical simulations revealing that, depending on the coupling strengths $(J,K_{1},K_{2})$, the system described by Eq.~\eqref{model} converges to five distinct long-term stationary states: `async', `thick phase wave', `thin phase wave', `spatially coherent', and `sync' states. In addition, three nonstationary states are observed: `gas state', `nonstationary thick phase wave', and `nonstationary thin phase wave'. Notably, two of these states, the stationary spatially coherent state and the nonstationary gas state, arise specifically due to the inclusion of higher-order interactions and do not appear when only pairwise interactions are present among the swarmalators \cite{o2024solvable}.

\par In the following, we provide a detailed analysis of these collective states. To enhance the clarity of these descriptions, we invite the reader to visualize the accompanying Supplementary Movies, which visually depict the behavior of the system in each of these states. With this remark in mind we will now describe the above mentioned states:
\begin{figure}[hpt]
	\centerline
	{\includegraphics[scale=0.12]{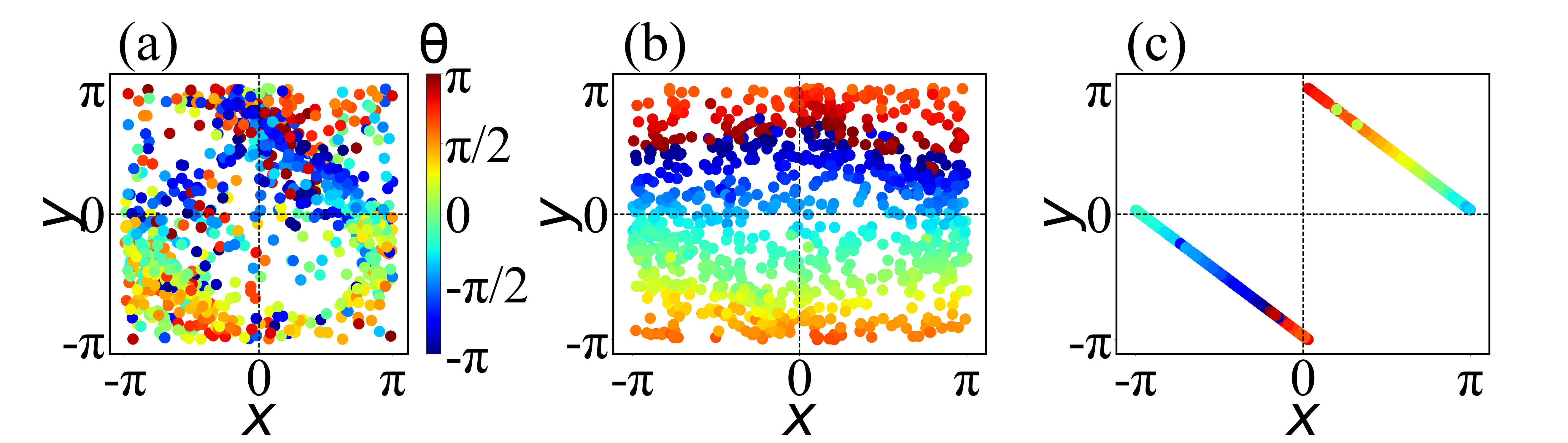}}
	\centerline
	{\includegraphics[scale=0.12]{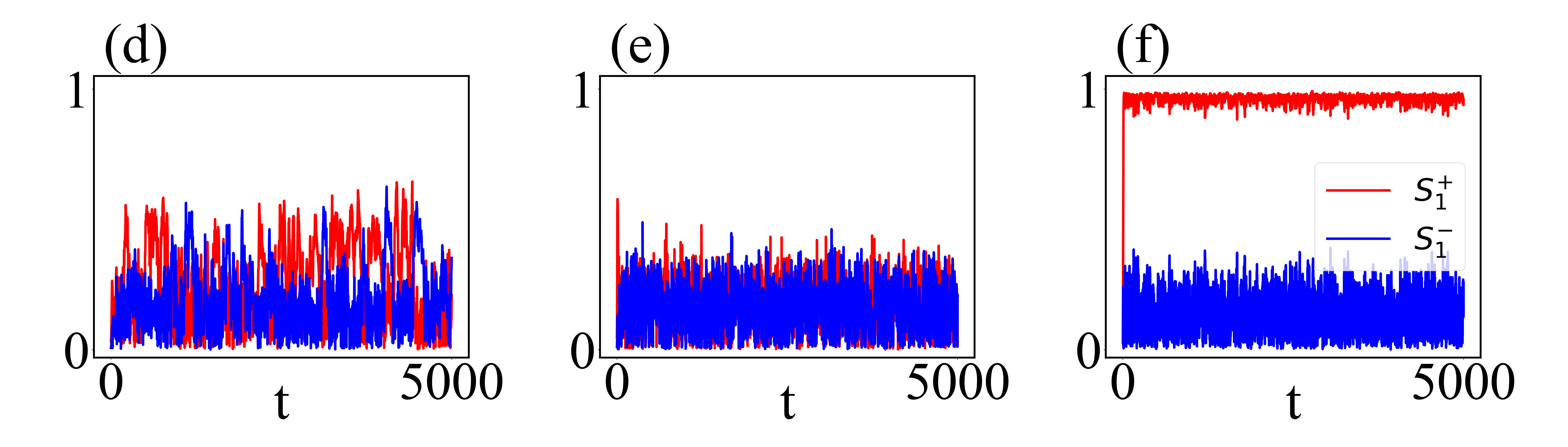}}
	\centerline
	{\includegraphics[scale=0.12]{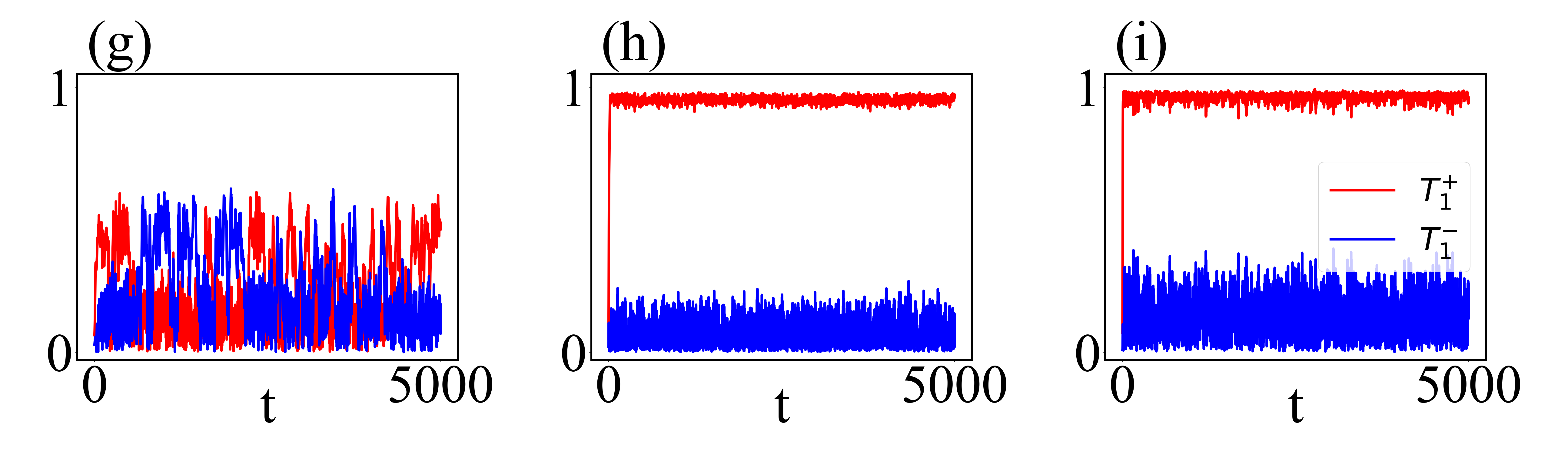}}
	\caption{{\bf Nonstationary states.} Top row: snapshots of the states in $(x,y)$ plane with colors indicating the phases. Middle row: order parameter $S_{1}^{\pm}$ as a function of time, and bottom row: order parameter $T_{1}^{\pm}$ as a function of time. (a, d, g) Gas state: $(J, K_1, K_2)=(1, -0.5, -4)$, (b, e, h) Thick phase wave: $(J, K_1, K_2)=(3, -1, -2)$, (c, f, i) Thin phase wave: $(J, K_1, K_2)=(4, -1, -1)$. The other parameters are $(N,T)=(1000,5000)$. }
	\label{active_states}
\end{figure}
\begin{itemize}
	\item \textit{Async state}: [Fig.~\ref{static_states}(a)] Swarmalators remain uniformly scattered with absolutely no correlation among the space and phase coordinates. As a result, the order parameters have the minimal value zero [Figs.~\ref{static_states}(f, k)].   
	
	\item \textit{Thick phase wave}: [Fig.~\ref{static_states}(b)] Here the swarmalators are uniformly distributed over the two-dimensional space $(x,y)$ with the phases being correlated with one of the space coordinates, i.e, $\theta_{i}= \pm x_{i}+c$ for all $i$ (or $\theta_{i}= \pm y_{i}+c$ for all $i$), where $c$ is an arbitrary constant that stems from the symmetry in the system dynamics and each $\pm$ variation is equally likely to occur. As a result, when $\theta_{i}$ are correlated with $x_{i}$, the order parameter $S_{1}^{+}=1$ and all the other order parameters have the zero value. Conversely, when phases the correlated with only the $y$-coordinated, the order parameter $T_{1}^{+}$ becomes $1$ and all the other have zero value [Fig.~\ref{static_states}(g, l)].  
	
	\item \textit{Thin phase wave}: [Fig.~\ref{static_states}(c)] In this case all the space and phase coordinates are correlated with each other, i.e., $\theta_{i}= \pm x_{i}+c= \pm y_{i}+c$ for all $i$. Consequently, the order parameters $S_{1}^{+}$ and $T_{1}^{+}$ reach the maximum value one, while the other order parameters  $S_{1}^{-}=T_{1}^{-}=0$ [Fig.~\ref{static_states}(h, m)]. 
	
	\item \textit{Spatially coherent}: [Fig.~\ref{static_states}(d)] For repulsive higher-order interactions $(K_{2}<0)$ and depending on the pairwise phase and space couplings, the swarmalators settle into a stationary configuration where they no longer move in space. In this state, the swarmalators collapse to fully synchronized spatial positions that are separated by $\pi$ unit, i.e., $x_{i},y_{i}=c+n\pi$ $(n=0,1)$. For certain initial conditions, the swarmalators converge to a single position $(n = 0)$, while for others, they occupy two positions separated by $\pi$ units $(n = 1)$. However, despite this spatial synchronization, the phases of the swarmalators remain distributed around two different phases. Consequently, all the order parameters $S_{1}^{+},T_{1}^{+},S_{1}^{-},T_{1}^{-}$ take values samller than $1$. The phases of the swarmalators show a certain level of coherence around two distinct phases: most of the swarmalators are phase locked while others drift around them.   
	
	\item \textit{Sync state}: [Fig.~\ref{static_states}(e)] Here, the swarmalators ultimately converge to fully synchronized clusters that are separated by $\pi$ unit, i.e., $x_{i}=c_{1}+n\pi,y_{i}=c_{2}+n\pi, \theta_{i}=c_{3}+n\pi$ $(n=0,1)$. For certain initial conditions, the swarmalators converge to a single cluster $(n = 0)$, while for others, they form two clusters separated by $\pi$ units $(n = 1)$. The order parameters in this state take the maximal value $S_{1}^{+}=T_{1}^{+}=S_{1}^{-}=T_{1}^{-}=1$.
	
	\item \textit{Gas state}: [Fig.~\ref{active_states}(a)] In the regime of repulsive higher-order coupling, the swarmalators dynamically self-organize into a state where both their spatial and phase coordinates exhibit uniform movement. This dynamic behavior allows for the emergence of a wide range of phase and spatial configurations among the swarmalators. However, as time progresses, the swarmalators tend to form small, transient clusters, which they subsequently leave, leading to asynchronous behavior. This evolving pattern of clustering and dispersal is evident from the chaotic evolution of the order parameters, as shown in Figs.~\ref{active_states}(d, g). The chaotic fluctuations in these order parameters indicate that the system does not settle into a long-term, stable clustering configuration, but rather continues to undergo dynamic reorganization.    
	
	\item \textit{Nonstationary thick phase wave}: [Fig.~\ref{active_states}(b)] In this configuration, the swarmalators organize themselves similarly to the stationary thick phase wave state, where their phases exhibit a certain degree of correlation with one of the spatial coordinates. However, unlike the stationary state, the swarmalators are not fixed in space and continue to move. This is evident from the oscillatory behavior of the order parameters shown in Figs.~\ref{active_states}(e, h). Here, $T_{1}^{+}$ oscillates around 1, indicating phase-space correlation, while the other order parameters oscillate near zero.
	
	\item \textit{Nonstationary thin phase wave}: [Fig.~\ref{active_states}(c)] In this case, all the phase and space coordinates exhibit a significant level of correlation, while the swarmalators continue to move in space. This dynamic is reflected in the behavior of the order parameters: $S_{1}^{+}$ and $T_{1}^{+}$ oscillate close to $1$, indicating a degree of correlation between the phase and space coordinates. Meanwhile, the other two order parameters oscillate around zero (Figs.~\ref{active_states}(f, i)).
	
\end{itemize}
\begin{figure*}[hpt]
	\centerline
	{\includegraphics[scale=0.33]{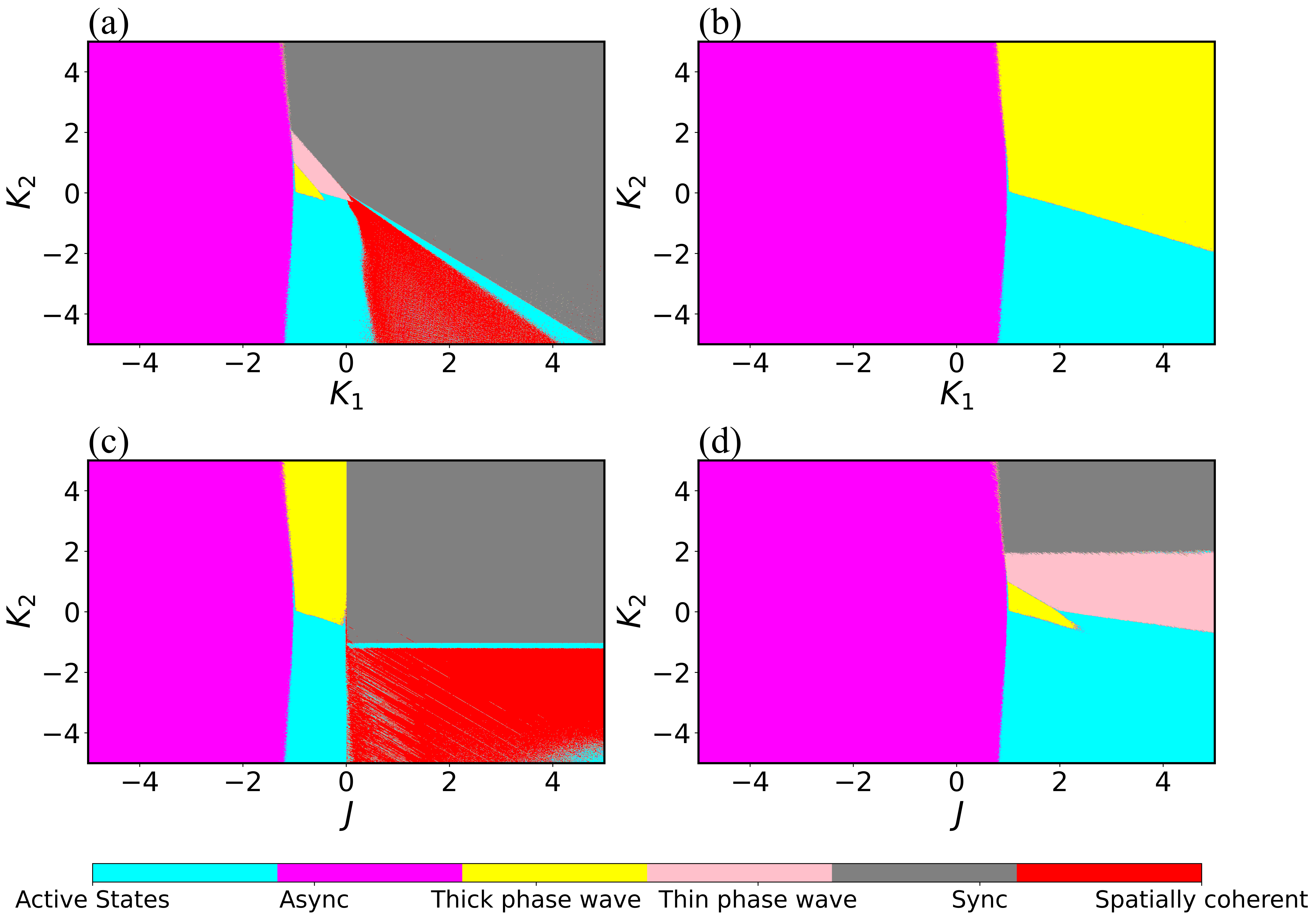}}
	\caption{{\bf Parameter regions for the emergence of different collective states} (a) $J=1$, (b) $J=-1$, (c) $K_{1}=1$, and (d) $K_{1}=-1$. The other parameters are $(N,T)=(1000,5000)$. The last $20\%$ data have been used to calculate the average of the order parameters and mean velocity $(V)$ that characterize the states. For the nonstationary states, $V >0$, while for the stationary ones $V \approx 0$. Further, the stationary states are characterized by using the values of order parameters $(S_{1}^{+}, T_{1}^{+})$.  Initial conditions are randomly drawn from $[-\pi,\pi]$ for $x_{i},y_{i}$, and $\theta_{i}$. It is important to note that with different initial configurations, the system may exhibit multistable behavior in certain parameter regions.}
	\label{parameter_space}
\end{figure*} 

\par Up to this point, we have explored how swarmalators self-organize into different collective states under specific coupling strengths. We now shift our focus to the dynamic emergence of these states as the coupling strengths are varied. Figure \ref{parameter_space} visually captures this phenomenon and the transitions between states by integrating Eq.~\eqref{model}, with initial positions and phases randomly selected from the interval $[-\pi, \pi]$. The results are organized into four distinct cases based on the signs of the spatial and phase (pairwise) couplings. We begin by examining the regimes of attractive and repulsive spatial couplings to identify where different states arise within the $(K_{1},K_{2})$ parameter plane. By rescaling time, we set $J=1$ for attractive spatial interactions and $J=-1$ for repulsive ones. Next, we explore the effects of attractive and repulsive pairwise phase couplings, mapping the occurrence of different states in the $(J,K_{2})$ parameter plane. Similarly, by rescaling time, we set $K_{1}=1$ and $K_{1}=-1$ to represent attractive and repulsive phase interactions, respectively. These findings offer deeper insights into how higher-order interactions shape the emergence of distinct collective states. In Fig.~\ref{parameter_space}, we first categorize the states into stationary and nonstationary ones by observing the mean velocity $V$, defined as: 
\begin{equation}
    \begin{array}{l}
         V=\Bigg\langle \dfrac{1}{N} \sum\limits_{j=1}^{N} \sqrt{\dot{x}_{j}^{2}+\dot{y}_{j}^{2}+\dot{\theta}_{j}^{2}}\Bigg\rangle_{t}, 
    \end{array}
\end{equation}
where $\langle \cdots \rangle_t$ represents the time average, taken after discarding an initial transient. If $V > 0$ the system lies in an nonstationary (active) state, while stationary states are characterized by $V \approx 0$. The stationary states are further distinguished by their order parameter values and color-coded accordingly. The async state is characterized by $(S_1^+, S_1^-, T_1^+, T_1^-) \approx (0, 0, 0, 0)$. For the thick phase wave, we observe $(S_1^+, S_1^-, T_1^+, T_1^-) \approx (1, 0, 0, 0)$ or $(0, 0, 1, 0)$, and for the thin phase wave, $(S_1^+, S_1^-, T_1^+, T_1^-) \approx (1, 0, 1, 0)$. The sync state is represented by $(S_1^+, S_1^-, T_1^+, T_1^-) \approx (1, 1, 1, 1)$, while the spatially coherent state lies in the range $0 < S_1^{\pm}<1$ and $0< T_1^{\pm} < 1$. Below, we discuss the results for the four distinct cases. It is important to note that the states observed in Fig.~\ref{parameter_space} emerge when the initial positions and phases of the swarmalators are chosen randomly within a cube of length $2\pi$, i.e., each value is randomly drawn from the interval $[-\pi,\pi]$. For other initial configurations, we observe that multiple states can coexist for the same set of parameters, demonstrating multistable behavior (which will be discussed later).  

\par (i) In Figure \ref{parameter_space}(a), we observe the emergence of collective states in the $(K_1, K_2)$ parameter plane for attractive spatial coupling ($J = 1$). When higher-order interactions are introduced, i.e., $K_2 > 0$, the system passes through four out of five possible stationary states in a small range of values of $K_1$. Indeed as $K_1$ increases from negative to positive, the system follows the sequence: async $\rightarrow$ thick phase wave $\rightarrow$ thin phase wave $\rightarrow$ sync state. With a further increase in $K_2$, the system undergoes a direct transition from async to sync state as $K_1$ increases, bypassing the intermediate thick and thin phase waves seen at lower $K_2$. When $K_2 < 0$ (repulsive higher-order coupling), the system transitions from the async state to the sync state through intermediate nonstationary and stationary spatially coherent states.   

\par (ii) Figure \ref{parameter_space}(b) illustrates the system behavior when a fixed repulsive spatial coupling $J = -1$ is introduced, while varying the coupling parameters $K_1$ and $K_2$. For $K_2 > 0$, the system passes from the async state to the thick phase wave state as $K_1$ moves from negative to positive. Notably, in this regime, the system does not achieve the thin phase wave or sync states (as we will demonstrate this analytically later on). For $K_2 < 0$, the system passes from async to thick phase wave through an intermediate nonstationary state, but only within a small range of $K_2$. For larger negative values of $K_2$, only the async and nonstationary states emerge as $K_1$ varies.    

\par (iii) Figure \ref{parameter_space}(c) depicts the emergence of collective states when a fixed attractive pairwise phase coupling of $K_1 = 1$ is introduced, while varying the spatial coupling $J$ and the higher-order coupling $K_2$. As $J$ increases from negative to positive, the system transitions from async to sync state via an intermediate thick phase wave for $K_2 > 0$, but the thin phase wave state does not emerge. For $K_2 < 0$, specifically beyond $K_2 = -1$, the sync state does not appear. Instead, the system moves from async to a spatially coherent state, passing through an intermediate nonstationary state.

\par (iv) Finally, the system behavior for a fixed repulsive phase coupling $K_{1}=-1$ is portrayed in Fig.~\ref{parameter_space}(d). Here, by varying $J$ and $K_{2}$, the system passes between async, thick phase wave, thin phase wave, sync, and nonstationary states. An intriguing observation, in this case, is the emergence of the sync state, which is surprising given that, in the pairwise model, the sync state never appears for repulsive phase coupling \cite{o2024solvable}. However, with higher-order interactions, we observe a significant region where the swarmalators settle into a sync state.
 
\subsection{Theoretical Analysis}
The aim of this section is to provide analytical support to prove the existence and the stability of the stationary states depending on the parameters $J$, $K_1$ and $K_2$. The nonstationary states, because of their very nature, challenge our theoretical study and thus we will limit our analysis to the stationary states.

\subsubsection{Analysis of async state}
To analyze the async state, we adopt the continuum limit approach (similar to the analysis presented in Ref.~\cite{o2024solvable}), where the incoherent state of the swarmalators is given by the density function $\rho(x,y,\theta)=\dfrac{1}{(2\pi)^3}$. By introducing a perturbation around this async state, we find that it loses its stability when $J+K_{1}=0$. Notably, this result indicates that the transition from the async state to other states is independent on the higher-order coupling, i.e., it holds true for any value of $K_{2}$.

\subsubsection{Analysis of sync state}
In the sync state, the system converges to a fixed point $(x_i,y_i,\theta_i)=(c_1, c_2, c_3)$, for all $i$, where the constants $c_{j}$ ($j=1,2,3$) are determined by the initial conditions. For simplicity, we consider the case of a single cluster, although one could also choose a configuration where clusters are separated by $\pi$. The stability conditions remain the same for both configurations. By linearizing the governing equation \eqref{model} around this fixed point, we obtain the eigenvalues of the associated Jacobian: $\lambda_{1}=0$, $\lambda_{2}=-J$, and $\lambda_{3}=-2(K_{1}+K_{2})$ with multiplicities of $3$, $(2N-2)$, and $(N-1)$, respectively. From the latter, we can deduce that the sync state is stable when $J>0$ and $K_{1}+K_{2}>0$. This tells us that the sync state can still emerge even when the pairwise phase interaction is repulsive  ($K_{1}<0$), as long as the higher-order phase coupling is attractive and its magnitude is greater than that of the pairwise coupling, i.e., $|K_{2}|>|K_{1}|$, with the spatial interactions remaining attractive (as depicted in Fig.~\ref{parameter_space}(d)).  

\par The stability conditions of the sync and async states together reveal an important insight: when $J>0$, both the sync and async states can coexist in the region where $|K_{2}|>|K_{1}|$, with $K_{1}<0$ and $K_{2}>0$. This indicates that the system exhibits bistability between the sync and async states within this region. In other words, depending on the initial conditions, the swarmalators may either converge to the sync state or remain in the async state. An instance of these bistability behaviors is illustrated in Appendix~\ref{bistabilities} (1).

\subsubsection{Analysis of thin phase wave state}
Since the phase and space coordinates of the swarmalators exhibit a perfect correlation, the fixed points in the thin phase wave state take the form
\begin{equation}
    \begin{array}{l}
         x_{i}=\dfrac{2\pi}{N}(i-1)+C_{1}, \\
         y_{i}=\dfrac{2\pi}{N}(i-1)+C_{2}, \\
         \theta_{i}=\dfrac{2\pi}{N}(i-1)+C_{3},
    \end{array}
\end{equation}
for $i=1,2,\dots,N$. By linearizing the system around the fixed point and evaluating the Jacobian, similar to the approach in Ref.~\cite{o2022collective}, we obtain the following eigenvalues,
\begin{equation}
    \begin{array}{l}
         \lambda_{1}=0, \lambda_{2}=-\dfrac{J}{4}, \lambda_{3}= -\dfrac{J}{2},  \lambda_{4}=\dfrac{1}{2}(-J-2K_1-2K_2), \\
         \lambda_{5,6}=\dfrac{1}{8} (-J-2K_1-4K_2 \pm \sqrt{J^2 +68 J K_1 +4k_1^2+ 40JK_2+16K_1K_2+16K_2^2}),
    \end{array}
\end{equation}
with multiplicities $N$, $2$, $(N-3)$, $(N-3)$, $2$, and $2$, respectively. The zero eigenvalue does not contribute to the stability of the state, as it stems from the rotational invariance of the system. The thin phase wave becomes stable when the real part of the remaining eigenvalues assumes negative value as we vary the coupling parameters. This eventually yields to the stability region determined by
\begin{equation}
	\begin{array}{l}
		J>0, 2(K_{1} + K_2) > -J,
		2K_1 +K_2<0,\; \text{and}\; 2(K_1 + 2 K_2) >- J.
	\end{array}
\end{equation}
An immediate conclusion is that the thin phase wave does not emerge when spatial interactions are repulsive (as portrayed in Fig.~\ref{parameter_space}(b)). 

\par Furthermore, the stability conditions of the async, sync and thin phase wave states altogether reveal an interesting phenomenon: the possibility of bistability. This means that the system can support both the thin phase wave and the async or sync state simultaneously, with the actual state depending on the initial conditions of the system. This bistability specifically arises when the pairwise phase interaction is repulsive, i.e., $(K_{1}<0)$. Appendix~\ref{bistabilities} (2) illustrates this bistability nature in more detail.

\subsubsection{Analysis of thick phase wave}
The stability region for the thick phase wave is more elusive compared to the thin phase wave. The thick phase wave, characterized by the correlation between the phase coordinates and one of the spatial coordinates of the swarmalators, is believed to bifurcate from the asynchronous (async) state and eventually to transit toward the thin phase wave state \cite{o2024solvable}. This suggests that the thick phase wave should emerge between these two states in the parameter space. However, while numerical simulations (as shown in Fig.~\ref{parameter_space}) support this hypothesis in regions where all three states (async, thick phase wave, and thin phase wave) are present, they also reveal something surprising: the thick phase wave can emerge even when the thin phase wave does not (Fig.~\ref{parameter_space}(b)). This observation complicates the theoretical understanding of the thick phase wave stability, as it indicates that the conditions for the latter may extend beyond the transitions between async and thin phase waves.
Thus, while numerical simulations provide evidence for the existence of the thick phase wave, the challenge of deriving an exact stability region for this state remains unresolved. 

\subsubsection{Analysis of spatially coherent state}
We are unable to provide the exact stability condition of this state. However, we provide an approximate condition for this state to emerge in the parameter region of interest. As discussed previously, the swarmalators are spatially fully synchronized in one or two clusters. We consider the case of a single cluster without loss of generality. Therefore, the space coordinates of the swarmalators are given by $x_{i}=C_{1}, y_{i}=C_{2}$ for all $i$. Because the equations of motion only depend on the differences of the positions, the system is invariant for translations and thus, without loss of generality, we can assume the positions of the swarmalators to be given by $(x_i,y_i)=(0,0)$ for all $i$. Therefore, the order parameters eventually become 
$S_{m}^{\pm}=T_{m}^{\pm}=R_{m}$ $(m=1,2)$, where $R_{m}=\dfrac{1}{N} \sum\limits_{j=1}^{N}e^{im\theta_{j}}$. This claim is confirmed by the numerical result shown in Fig.~\ref{static_states}(i, n). Now, plugging the fixed point conditions in the governing equations gives us the following
\begin{equation}
	\begin{array}{l}
		\dfrac{2K_{1}}{N} \sum\limits_{j=1}^{N} \sin(\theta_{j}-\theta_{i}) + \dfrac{2K_{2}}{N^2} \sum\limits_{j,k=1}^{N} \sin(2\theta_{j}-\theta_{k}-\theta_{i})=0,
	\end{array}
\end{equation} 
which eventually simplifies into 
\begin{equation}
	\begin{array}{l}
		K_{1}+K_{2}R_{2}=0.
	\end{array}
\end{equation}
The phases of the swarmalators converge to exactly two distinct points when $R_{2}=1$, while for $R_{2}=0$, they remain uniformly distributed. Therefore, for the spatially coherent state $R_{2}$ must satisfies the relation $0<R_{2}<1$. This provides us the region of emergence of the spatially coherent state as $K_{1}>0$ and $0<-\dfrac{K_{1}}{K_{2}}<1$. Now, the spatial coordinates of the swarmalators can fully synchronize only when $J>0$. Thus, we ultimately obtain the emergence condition for this state as, 
\begin{equation}
	\begin{array}{l}
		0<-\dfrac{K_{1}}{K_{2}}<1, 
	\end{array}
\end{equation}
with $J,K_{1}>0$, and $K_{2}<0$. The region of parameters shown in Fig.~\ref{parameter_space} support the above result, indeed the region of emergence of spatially coherent state is defined by $K_2<-1$ and $J>0$, having fixed $K_1=1$. 

\par This completes our investigation with the identical swarmalators. In the following section, we will discuss the more realistic case with nonidentical swarmalators.

\section{Nonidentical swarmalators}
We hereby consider a population of nonidentical swarmalators whose free velocities ($u_i,v_i$) and natural frequencies ($w_i$) are drawn from Lorentzian distribution $g(z)={\Delta}/{\pi(z^2+\Delta^2)}$, with zero center and half-width $\Delta$. Numerical simulations reveal the emergence of the async, thick phase wave, thin phase wave and the sync states, in this setting as well. We also report the occurrence of the intermediate mixed states where the nonzero order parameters are unequal in values. The details of these states are further elaborated in Appendix \ref{ni_states}. In the following, we will mainly focus on the analysis of these collective behaviors.   

\par For the sake of simplicity and to simplify the mathematical analysis, we change coordinates and we deal with the ``sum-difference'' coordinates , namely $x_{\pm}=x \pm \theta$ and  $y_{\pm} = y \pm \theta$. In the thermodynamic limit $N\rightarrow \infty$, the density $\rho(u,v,w,x_{\pm},y_{\pm},t)$ obeys the continuity equation
\begin{equation}
	\dfrac{\partial \rho }{\partial t} + \nabla \cdot (v \rho) = 0. \label{cont}
\end{equation}
The equations for the derivatives in the new coordinates can be written as
\begin{subequations}
	\begin{align}
		v_{x_+} = &u_+ + \text{Im}\Big[H_+(W_1^+,W_2^+) e^{-i x_+} + H_-(W_1^-,W_2^-) e^{-i x_-} + G(Z_1^+,Z_2^+) e^{-i y_+} - G(Z_1^-,Z_2^-) e^{-i y_-} \Big],
	\end{align}
	\begin{align}
		v_{x_-} = &u_- + \text{Im}\Big[H_-(W_1^+,W_2^+) e^{-i x_+} + H_+(W_1^-,W_2^-) e^{-i x_-} - G(Z_1^+,Z_2^+) e^{-i y_+} + G(Z_1^-,Z_2^-) e^{-i y_-} \Big],
	\end{align}
	\begin{align}
		v_{y_+} = &v_+ + \text{Im}\Big[G(W_1^+,W_2^+) e^{-i x_+} - G(W_1^-,W_2^-) e^{-i x_-} + H_+(Z_1^+,Z_2^+) e^{-i y_+} + H_-(Z_1^-,Z_2^-) e^{-i y_-} \Big],
	\end{align}
	\begin{align}
		v_{y_-} = &v_- + \text{Im}\Big[-G(W_1^+,W_2^+) e^{-i x_+} + G(W_1^-,W_2^-) e^{-i x_-} + H_-(Z_1^+,Z_2^+) e^{-i y_+} + H_+(Z_1^-,Z_2^-) e^{-i y_-} \Big],
	\end{align}
\end{subequations}
where $u_{\pm}=u\pm w, v_{\pm} = v \pm w$, and the complex functions $H_{\pm}$ and $G$ are defined as
\begin{subequations}
\begin{align}
	H_{\pm}(z_1,z_2) &= \frac{J \pm K_1}{2} z_1 \pm \frac{K_2}{2} z_2 z_1^*,
\end{align}
\begin{align}
	G(z_1,z_2) &= \frac{K_1}{2} z_1 + \frac{K_2}{2} z_2 z_1^*.
\end{align}
\end{subequations}
The strategy is to derive expressions for the order parameters by using a generalized Ott-Antonsen (OA) ansatz \cite{ott2008low}. While the Kuramoto model is associated to dynamics on the unit circle $\theta \in \mathbb{S}^1$, our model corresponds to synchronization on $(x, y, \theta) \in \mathbb{S}^1 \times \mathbb{S}^1 \times \mathbb{S}^1$. Conveniently, we search for an OA ansatz in the form of a product of Poisson kernels. In the $(x_{\pm}, y_{\pm})$ coordinates, the latter is of the form
\begin{equation}
	\begin{array}{l}
		\rho(u,v,w,x_{\pm}, y_{\pm},t) = \frac{1}{16 \pi^4} g(u) g(v) g(w) \times \Big[ 1 + \sum_{n=0}^{\infty} \alpha_x^n e^{i n x_+} + \text{c.c.} \Big] \times [1 + \sum_{m=0}^{\infty} \beta_x^m e^{i m x_-} + \text{c.c.} ]  \\
	 \times \Big[ 1 + \sum_{l=0}^{\infty} \alpha_y^l e^{i l y_+} + \text{c.c.} \Big]  \times [1 + \sum_{p=0}^{\infty} \beta_y^p e^{i p y_-} + \text{c.c.} ],
\end{array}
\end{equation}
where ``c.c." denotes complex conjugate terms. We substitute this ansatz into the continuity equation~\eqref{cont} and then project onto $e^{i x_{\pm}}, e^{i y_{\pm}}$ to yield coupled complex differential equations for the Fourier modes associated to $n,m,l$, and $p$ in the submanifold $||\alpha_{x,y}|| = ||\beta_{x,y}|| = 1$. See Appendix~\ref{appendixb} for more details. The expressions for the rainbow order parameters become,
\begin{align}
	W_1^+ = S_1^+ e^{i \phi_1^+} = \int \alpha_{x}^*(u,v,w) g(u)g(v)g(w) du dv dw ,\label{op}
\end{align}
and similarly for the remaining order parameters. Next, the expressions of the order parameters are derived by using the amplitude equations. We use Eqs.~\eqref{alpha-beta} to find fixed point expressions for $\alpha_{x}, \beta_x$ etc. in the emerging states, and then substitute those into the integrals for $W_1^+, W_1^-$ etc. $\phi_1^{\pm}, \psi_1^{\pm}$ can be set to zero by going to a suitable frame. For simplicity, we consider a particular case where $\Delta_u=\Delta_v=\Delta_w=1$.

{\it Sync state}: We already know from the analysis performed in the case of the identical swarmalators, that in the sync state the value of the order parameters are nonzero and equal each other, i.e., $S_1^+=S_1^-=T_1^+=T_1^-=S\neq 0$. In case of the nonidentical swarmalators, we look for a solution of Eq.~\eqref{alpha-beta} that satisfies $\dot{\alpha}_x = \dot{\beta}_x = \dot{\alpha}_y = \dot{\beta}_y = 0$ \cite{yoon2022sync}. We find
\begin{equation}
	\alpha_x(u,w) = \mathcal{H} \left[\frac{u}{J}+ \frac{w}{2(K_1+K_2S^2)}\right],
\end{equation}
and similarly for the remaining variables, where we introduced the complex-valued function $\mathcal{H}(z) = - i z + \sqrt{1-z^2}$. We solve the integral in Eq.~\eqref{op} by using the Residue theorem which leads us to the equation of the order parameter
\begin{equation}
	S = \mathcal{H}^* \left[i \left(\frac{1}{J}+ \frac{1}{2(K_1+K_2S^2)}\right)\right]. 
\end{equation}
The above equation can be associated with a six degree polynomial of $S$, the real branch of which corresponds to the solution of the order parameter
\begin{equation}
	S = \sqrt{\frac{\pm \sqrt{\frac{(J K_1+(J-2) K_2)^2}{J^2}-4 K_2}-K_1+K_2}{2 K_2}-\frac{1}{J}},
	\label{sync_branches}
\end{equation}
where `$+$' and `$-$' sings correspond to the stable and unstable solutions when they exist. The expression in Equation \eqref{sync_branches} indicates that, depending on the coupling strengths, both stable and unstable synchronization branches coexist when $K_{2}>K_{2}^{c}=\dfrac{J^2}{(J-2)^2}$. This implies that for $K_{2}>K_{2}^{c}$, two distinct transition scenarios are possible: a ``forward transition," where the order parameter bifurcates from $0$ to $S$, and a "backward transition," where the order parameter bifurcates from $S$ back to $0$. When $K_{2}<K_{2}^{c}$, the forward and backward transition points coincide, leading to a smooth transition in the system. For $K_{2}>K_{2}^{c}$, the two transition points differ, signaling the occurrence of an abrupt transition. By equating the constant term in the polynomial on the right-hand side of Eq.~\eqref{sync_branches} to zero, we can determine the boundary of the region where $S$ bifurcates from zero, representing the forward transition boundary. This boundary is given by $K_1=\frac{J}{J-2}$. In the backward transition, both stable and unstable branches coexist within a specific range of parameter values. When the backward critical coupling condition is met, these branches collide and annihilate each other, marking the backward transition boundary. The latter is obtained by equating the solutions of the stable and unstable branches, by yielding $K_1 = \left(\frac{2}{J}-1\right) K_2+2 \sqrt{K_2}$.

{\it Thin phase wave state}: In the case under consideration the order parameters satisfy $(S_1^+,S_1^-,T_1^+,T_1^-)=(S,0,S,0)$. We are thus interested in determining solutions of Eq.~\eqref{alpha-beta} satisfying $\dot{\alpha}_x = \dot{\alpha}_y = 0$ and $\dot{\beta}_{x,y} \ne 0$. Repeating the analysis performed in the sync state, we can obtain the following expression for the nonzero order parameter:
\begin{equation}
	S=\frac{1}{2} \sqrt{\frac{\pm \sqrt{A}-J-2 K_1}{K_2}-\frac{8}{J}+2},
\end{equation}
where 
\begin{equation*}
    A=\frac{4 (J-4)^2 K_2^2}{J^2}+\frac{8 (J-4) K_1 K_2}{J}+(J+2 K_1)^2+4 (J-12) K_2\, ,
\end{equation*}
let us observe that the `$+$' and `$-$' signs correspond to the stable and unstable solutions. For $K_{2}> \frac{2J^2}{(J-4)^2}$, both the stable and unstable solution coexist within specific range of coupling strengths. The forward boundary of this state is found to be $J^2+2 J K_1-8 J-8 K_1=0$, and the backward transition threshold is given by $K_1 = \frac{4 K_2}{J}-\frac{J}{2}-K_2+2 \sqrt{2 K_2}$.

{\it Thick phase wave state}: In the thick phase wave the order parameters are of the form $(S_1^+,S_1^-,T_1^+,T_1^-)=(S,0,0,0)$, we hence look for solutions of the Fourier amplitudes such that $\dot{\alpha}_x = 0$ and $\dot{\alpha}_y, \dot{\beta}_{x,y} \ne 0$. The same analysis as in the previous two cases gives us the order parameter
\begin{equation}
	S=\sqrt{\frac{\pm \sqrt{(J+K_1+K_2)^2-32 K_2}-J-K_1+K_2}{2 K_2}},
\end{equation}
with `$+$' and `$-$' signs corresponding to the stable and unstable solutions. Here, for $K_{2}>8$, both the stable and unstable solution branches emerge for specific coupling regimes. The forward transition boundary is $J+K_1=8$, while the backward transition point is given by $K_1 = -J-K_2+4 \sqrt{2 K_2}$.

{\it Async state}: The async state is defined by $(S_1^+,S_1^-,T_1^+,T_1^-)=(0,0,0,0)$, we thus determine its stability by considering a small perturbation about the steady state solutions
\begin{subequations}
	\begin{align}
		\alpha_{x,0}(u,v,w,t) &= \exp \left[-i (u+w) t\right],
	\end{align}
	\begin{align}
		\beta_{x,0}(u,v,w,t) &= \exp \left[-i (u-w) t\right],
	\end{align}
	\begin{align}
		\alpha_{y,0}(u,v,w,t) &= \exp \left[-i (v+w) t\right],
	\end{align}
	\begin{align}
		\beta_{y,0}(u,v,w,t) &= \exp \left[-i (v-w) t\right].
	\end{align}
\end{subequations}
Given $\epsilon>0$ a small parameter, we can write the perturbation at the first order in the form $\alpha_{x}(u,v,w,t) = \alpha_{x,0}(u,v,w,t) + \epsilon \alpha_{x,1}(u,v,w,t)$, then the perturbed order parameters will result to be $W_1^{+(1)}  = \int \alpha_{x,1}^*(u,v,w) g(u)g(v)g(w) du dv dw$. It is easy to determine that the perturbed order parameters decay when $J+K_{1}<8$ which coincides with the forward transition boundary of thick phase wave~\cite{yoon2022sync,anwar2024collective}.

{\it Mixed state}: We were unable to resolve the mixed states because they are not associated to any fixed points for the amplitudes $\alpha_x,\beta_x,\alpha_y,\beta_y$. Let us observe that these states are somewhat unusual and are not seen in models like the Kuramoto one, or in the case of identical swarmalators. Without any formal analysis, we cannot precisely determine the properties of the mixed states, the latter could potentially arise with other states (i.e., to be bistable) or even represent a long transient phase.


\begin{figure}[hpt]
	\centerline{\includegraphics[scale=0.5]{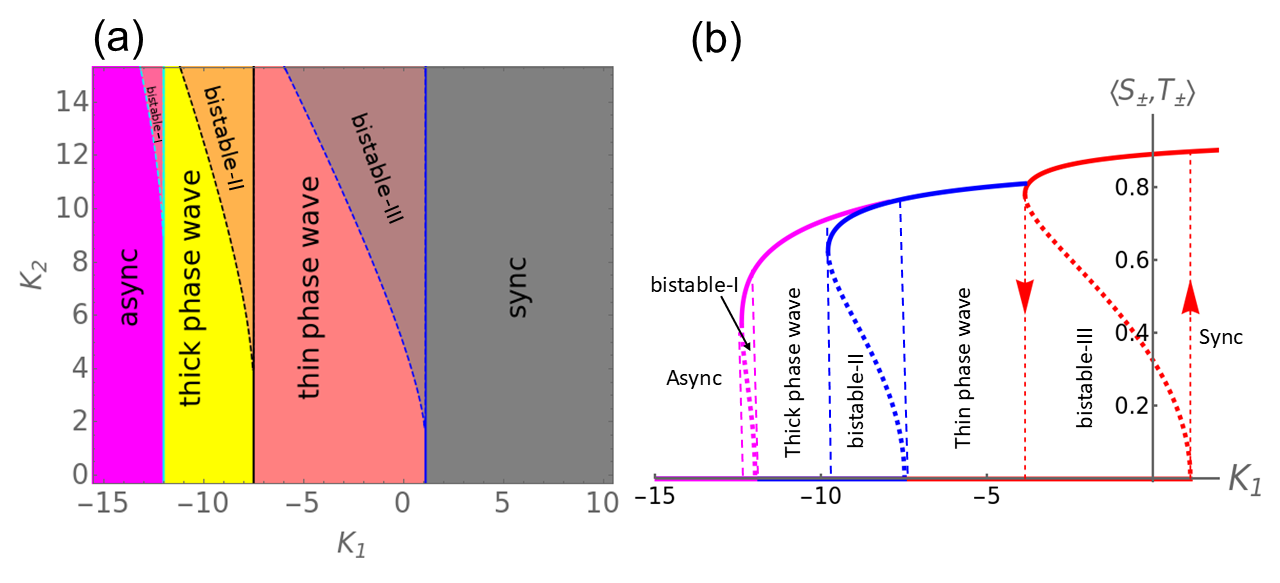}}	
	\caption{(a) {\bf Phase transition scenario in the $K_1$-$K_2$ plane for nonidentical swarmalators.} The spatial coupling strength $J$ is fixed here ($J=20$). We observe bistability between the collective states which is clearly visible for large $K_2$. The transitions are abrupt, and the solid and dashed lines correspond to the points of forward and backward transitions, respectively. We also notice that, when higher-order coupling is absent (i.e., $K_2=0$) or sufficiently small, the nature of the transition is smooth. (b) Dependence of the order parameters with respect to $K_1$ for Fixed $J=20$ and $K_2=12$. The solid curves represent the stable branches of the solution associated to the specific value of the order parameter, while the dotted curves denote unstable branches. The vertical dotted lines mark the critical values of $K_1$ where forward (right) and backward (left) transitions occur. The system exhibits three distinct bistable regions for this parameter set.}
	\label{nonid1}
\end{figure}

\par The above analysis has revealed that we can explicitly work out the expressions of the order parameters in all the collective states except the mixed state and also their stable/unstable boundaries. This fact highlights the solvability of the model despite its complexity. Another fascinating point is the occurrence of abrupt transition between the collective states when one of the system parameters is varied. Higher-order interaction is responsible for this discontinuous phase transition. In presence of sufficiently strong non-pairwise phase coupling, swarmalators overcome the effect of heterogeneous frequencies and also dominate the influence of pairwise coupling. For instance, we depict the interplay between the pairwise and non-pairwise coupling strengths when $J$ is fixed to $20$ (see Fig.~\ref{nonid1}(a)). Figure~\ref{nonid1}(a) shows the phase transition phenomena in the $K_1$-$K_2$ plane; by increasing $K_1$, we observe that the system goes through the transitions: async (magenta region) $\rightarrow$ thick phase wave (yellow region) $\rightarrow$ thin phase wave (pink region) $\rightarrow$ the sync state (gray region). The forward, i.e., once $K_1$ increases, boundaries between these regions are denoted by the cyan, black, and blue solid lines, respectively. Their positions are analytically obtained by substituting $J=20$ in the expressions for the boundaries calculated previously; let us observe that they are independent from the higher-order phase coupling $K_{2}$. We also observe in Fig.~\ref{nonid1}(a) that there are regions of bistability (resulting from abrupt phase transitions) between  successive states, which are especially prominent for large values of the non-pairwise coupling strength $K_2$ (as discussed before the bistabilities emerge beyond a critical value of $K_{2}$). The boundaries of those backward transition, i.e., by decreasing $K_1$, are denoted by blue, black, and cyan dashed lines, respectively from sync to thin phase wave, from thin phase wave to thick phase wave, and from thick phase wave to async. To provide a clearer description of these abrupt transitions and associated bistability regions, we report in Fig.~\ref{nonid1}(b) the variation of order parameters with respect to $K_1$ for fixed $J=20$ and $K_2=12$. The order parameters exhibit three distinct abrupt transitions, giving rise to three bistable regions: bistable-I: between the asynchronous state and the thick phase wave state, bistable-II: between the thin and thick phase wave states and bistable-III: between the thin phase wave and synchronized states, respectively. These findings emphasize the crucial role of higher-order interactions in shaping the collective dynamics of nonidentical swarmalators.

\section{Discussions}
Our study reveals that higher-order interaction brings fascinating collective states in the swarmalator system. The noticeable result is the presence of multiple bistable regions between the emerging dynamical states. In the case of identical swarmalators, we use fixed point analysis to derive the boundaries separating the different behaviors. The model involving nonidentical swarmalators is solved by using an Ott-Antonsen ansatz by assuming a product of Poisson kernels. The expressions of the nonzero order parameters in the different collective states are analytically derived by using  Fourier modes and from the integrals involving the order parameters. By summarizing our techniques, we get a polynomial equation for the order parameter, whose constant term determines the boundary where it bifurcates from zero. The real solutions of it correspond to the forward and backward transition curves. The region of bistability is then calculated by inspecting the existence condition of the unstable branch.

In conclusion, we have made significant progress in understanding the dynamics of two-dimensional swarmalators (under periodic boundary conditions) with higher-order phase coupling. Our analysis has unveiled the crucial role of non-pairwise interactions in shaping the collective behavior of the system, including abrupt phase transitions and the emergence of bistability between different collective states. However, several open questions remain. Despite being a stationary state, the thick phase wave could not be fully analyzed, particularly in terms of its stability. Recent techniques to determine the stability of compactly supported states (such as the thick phase wave) have been introduced for one-dimensional swarmalator models \cite{o2024stability}. We hope that future studies will extend these methods to higher-dimensional models to determine the stability conditions for the thick phase wave state in two-dimensional swarmalator systems. Additionally, in the case of identical swarmalators, the gas state exhibits irregular, potentially chaotic, dynamics that we could not fully characterize. Further investigations are required to clarify the nature of these chaotic solutions. Another challenging open problem is the description of mixed states observed in heterogeneous swarmalators, which could not be solved by using the techniques developed in this work. The intricate dynamics of these states demand novel analytical approaches. Future studies could also investigate higher-order interactions not only in phase dynamics but also in the spatial coordinates $(x, y)$. This would provide a more comprehensive understanding of the impact of non-pairwise interactions on swarmalator systems and potentially unveil new forms of collective behavior in complex systems.

\appendix

\section{Bistability analysis} \label{bistabilities}
Here, we illustrate some of the bistability behavior between different collective states exhibited by the system with identical swarmalators. 
\subsection{Bistability between async and sync state}
\begin{figure}[hpt]
    \centerline
    {\includegraphics[scale=0.25]{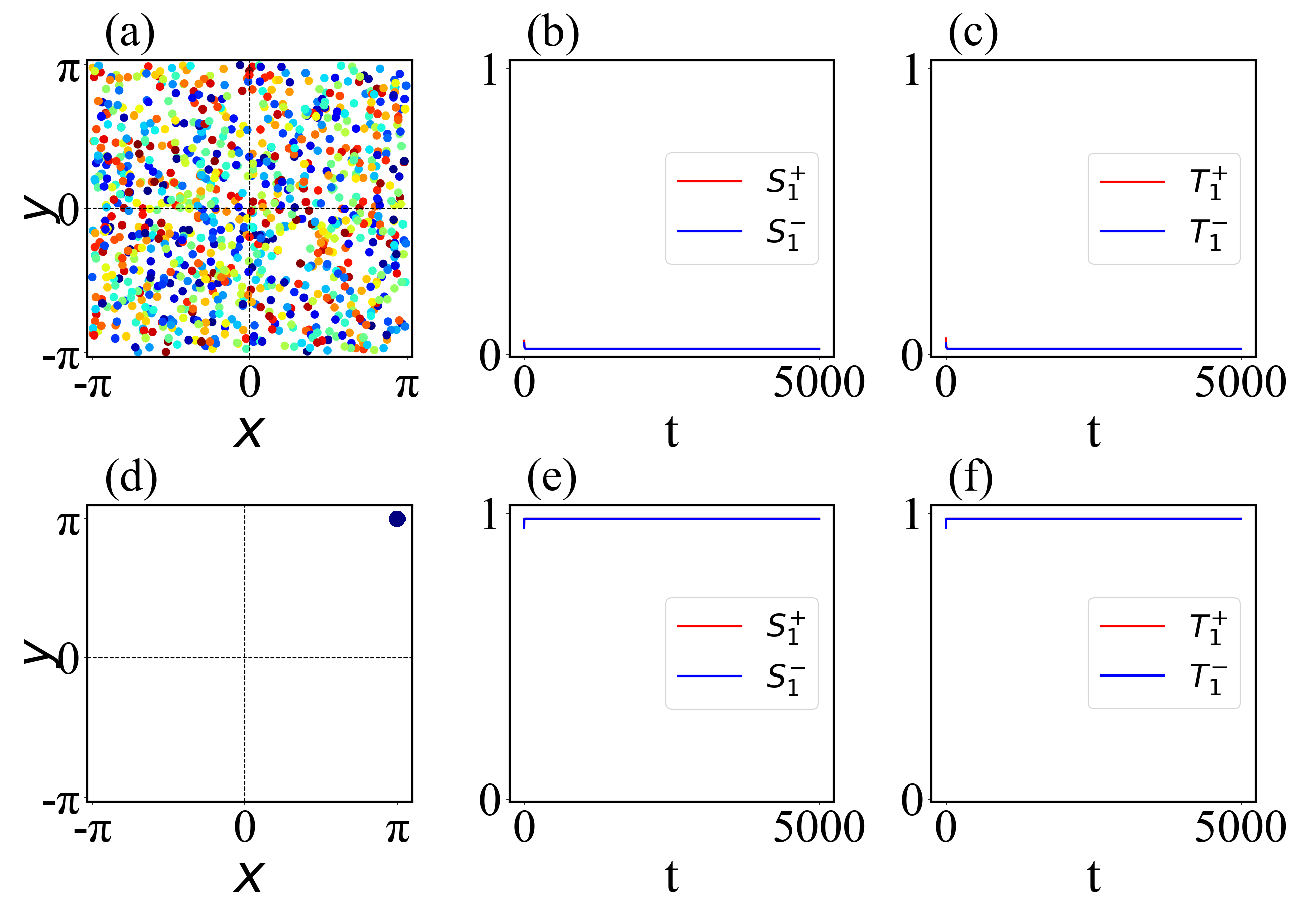}}
    \caption{ Bistability between async and sync state for $(J,K_1,K_2)=(1,-2,3)$. The Upper and lower rows correspond to the snapshots and time evolution of the order parameters for the async and sync states, respectively.}
    \label{async_sync_bistable}
\end{figure}

As discussed in the main text, for $J>0$, and $K_{1}<0$, both the sync and async state can coexist in the region of attractive higher-order coupling with $|K_{2}|>|K_{1}|$. Specifically, in this parameter region, the async state emerges when the initial positions and phases of the swarmalators are randomly distributed within a cube of length $2\pi$ units, i.e., $x_{i},y_{i},\theta_{i} \in [\pi,\pi]$ (as shown in Fig.~\ref{parameter_space}. We thus draw a point $(J,K_1,K_2)=(1,-2,3)$ from this parameter region. The upper panel of Fig.~\ref{async_sync_bistable} shows that with initial conditions drawn from the cube of length $2\pi$, swarmalators remain in the async state. In contrast, in the lower panel of Fig.~\ref{async_sync_bistable}, the initial conditions are drawn from a much smaller cube, such that the swarmalators start out very close to each other. Under these initial conditions, the swarmalators organize into the sync state. The accompanying plots of the order parameters further confirm this bistability.       

\subsection{Bistability between thin phase wave and async (sync) state}

\begin{figure}[hpt]
	\centerline
	{\includegraphics[scale=0.25]{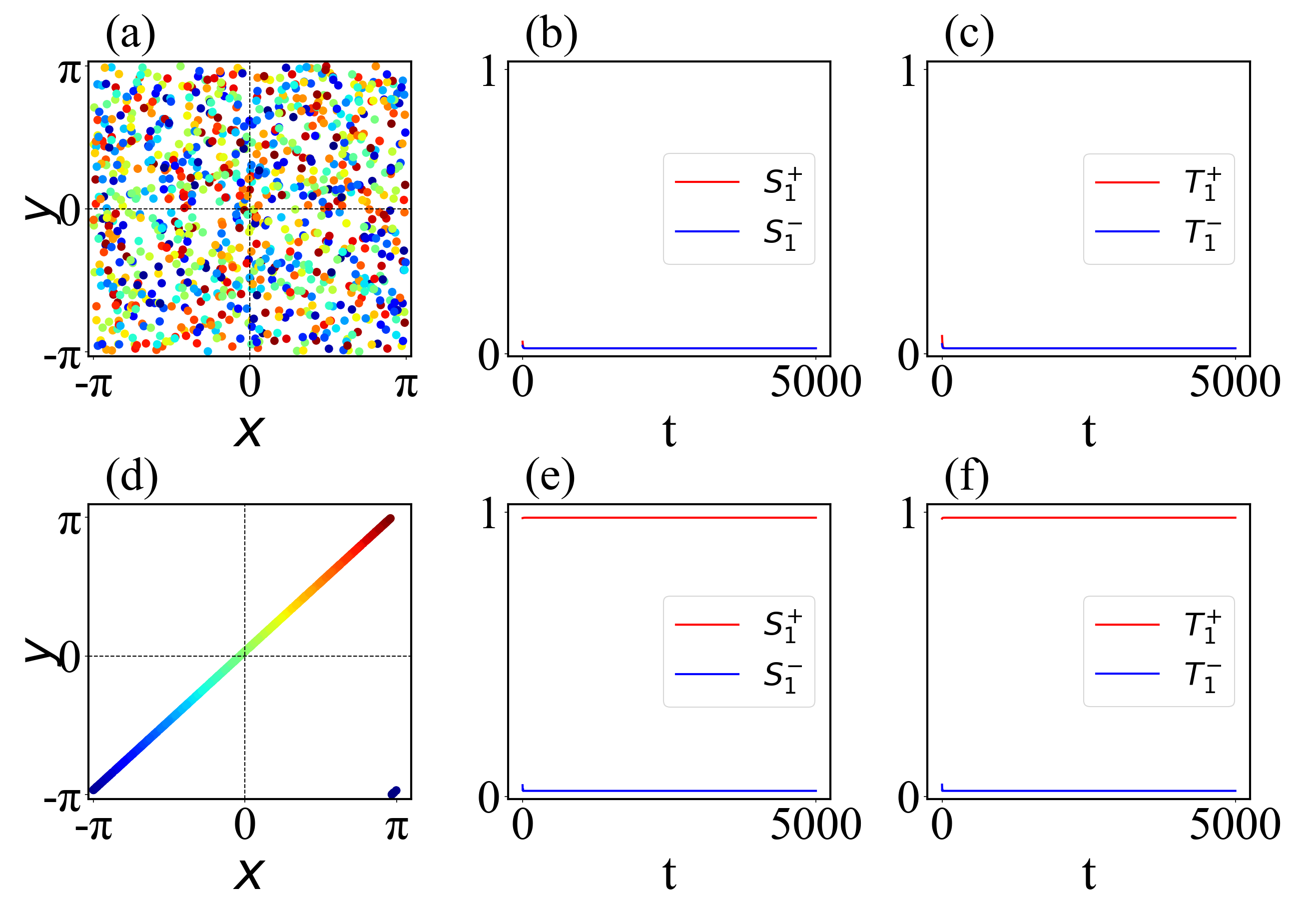}}
	\caption{ Bistability between async and thin phase wave state for $(J,K_1,K_2)=(0.5,-1,0.8)$. The Upper and lower rows correspond to the snapshots and time evolution of the order parameters for the async and thin phase wave states, respectively.}
	\label{async_thin_bistable}
\end{figure}

For $J>0$ and $K_{1}<0$, the system exhibits bistability between the async and thin phase wave state, as well as between the thin phase wave and sync state, depending on the choice of initial conditions. Figure~\ref{async_thin_bistable} displays the bistability between the async (upper panel) and thin phase wave (lower panel) state for $(J,K_1,K_2)=(0.5,-1,0.8)$. Here, the swarmalators exhibit async state when the initial phases and positions are chosen from the cube of length $2\pi$, whereas the thin phase wave state emerges when the initial conditions are chosen from the close neighborhood of the fixed points of the thin phase wave state. 

\par
\begin{figure}[hpt]
    \centerline
    {\includegraphics[scale=0.25]{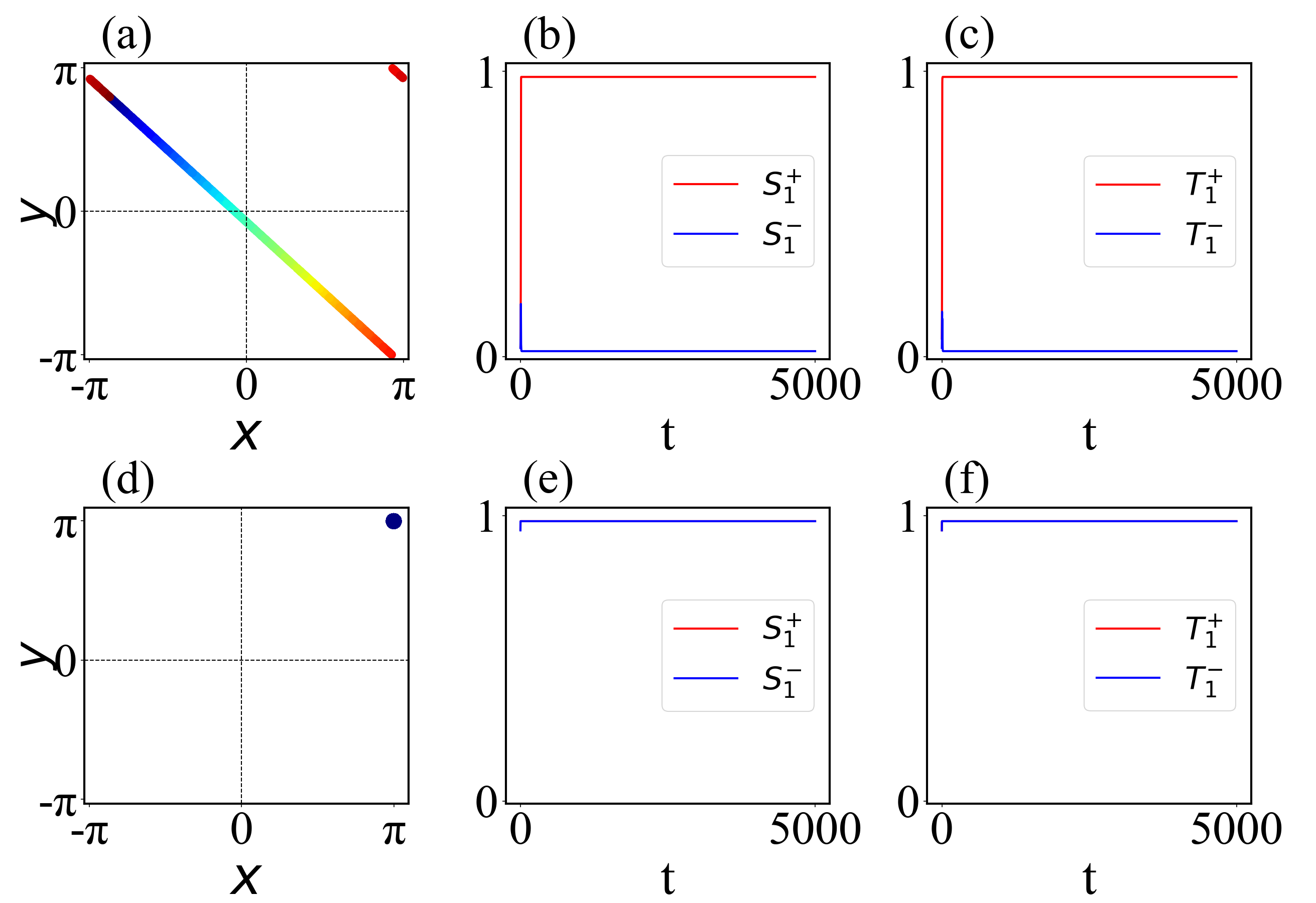}}
    \caption{Bistability between thin phase wave and sync state for $(J,K_1,K_2)=(3,-1,1.2)$. The Upper and lower rows correspond to the snapshots and time evolution of the order parameters for the thin phase wave and sync states, respectively.}
    \label{thin_sync_bistable}
\end{figure}

 Figure \ref{thin_sync_bistable} shows the coexistence of the thin phase wave (upper panel) and the sync state (lower panel) for $(J,K_1,K_2)=(3,-1,1.2)$. Here, the thin phase wave emerges for initial conditions drawn for the larger cube of length $2\pi$, while the sync state occurs when the initials are chosen from a small cube with swarmalators starting from very close to each other.     

\subsection{Bistability between thick phase wave and async (thin phase wave) state}
\begin{figure}[hpt]
	\centerline
	{\includegraphics[scale=0.25]{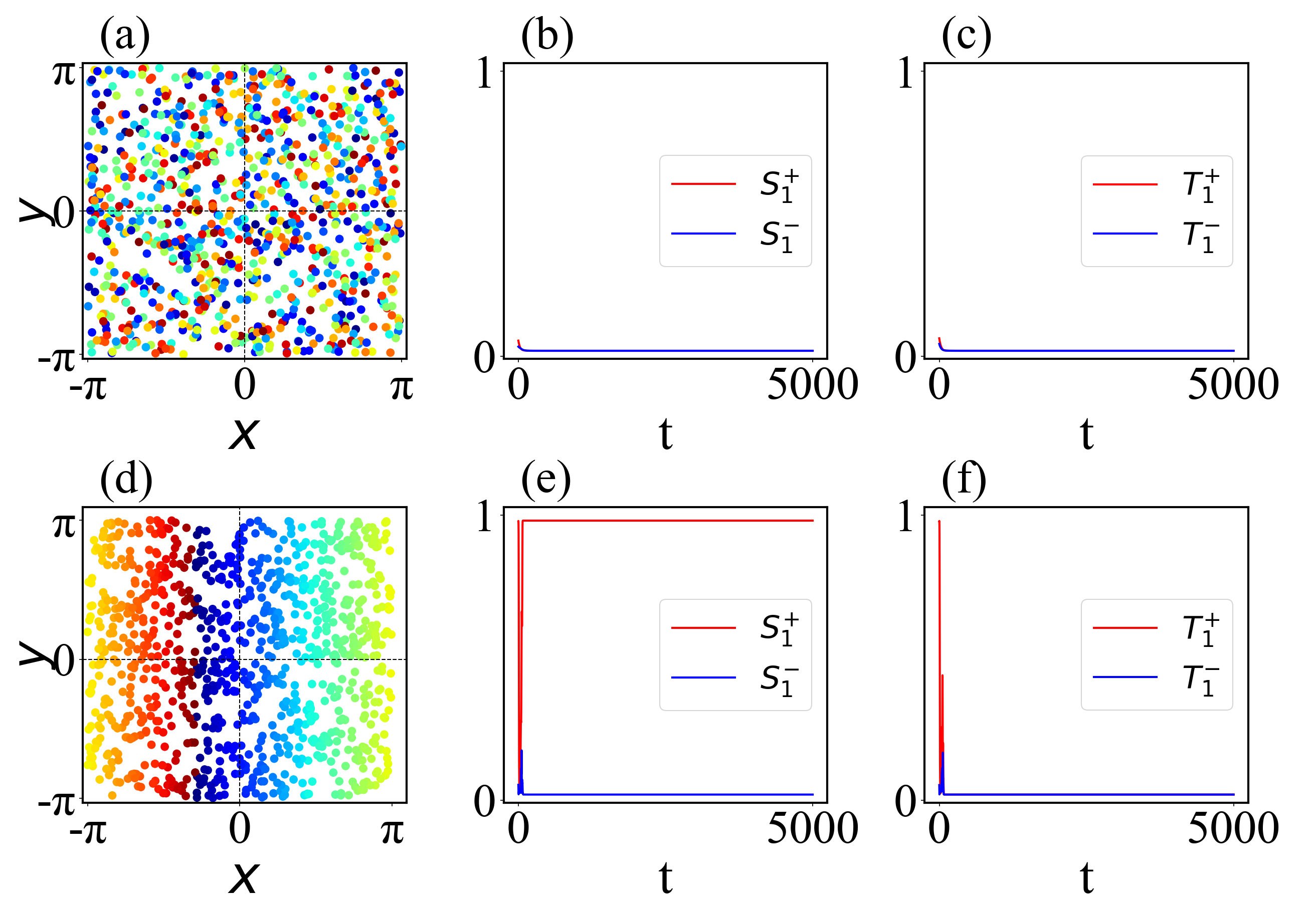}}
	\caption{ Bistability between async and thick phase wave state for $(J,K_1,K_2)=(-1,0.8,2)$. The Upper and lower rows correspond to the snapshots and time evolution of the order parameters for the async and thick phase wave states, respectively.} 
	\label{async_thick_bistable}
\end{figure}

The system further exhibits bistability between the thick phase wave and async state, as well as between the thick and thin phase wave state. Figure \ref{async_thick_bistable} illustrates the coexistence of the async and thick phase wave states for the parameter point $(J,K_1,K_2)=(-1,0.8,2)$. when the initial conditions are drawn randomly from the cube of length $2\pi$, the swarmalators settle into the async state (upper panel), while for the initial condition chosen from the close neighborhood of the fixed point of the thick phase wave, the swarmalators organize themselves to form thick phase wave state (lower panel). 
\par
\begin{figure}[hpt]
    \centerline
    {\includegraphics[scale=0.25]{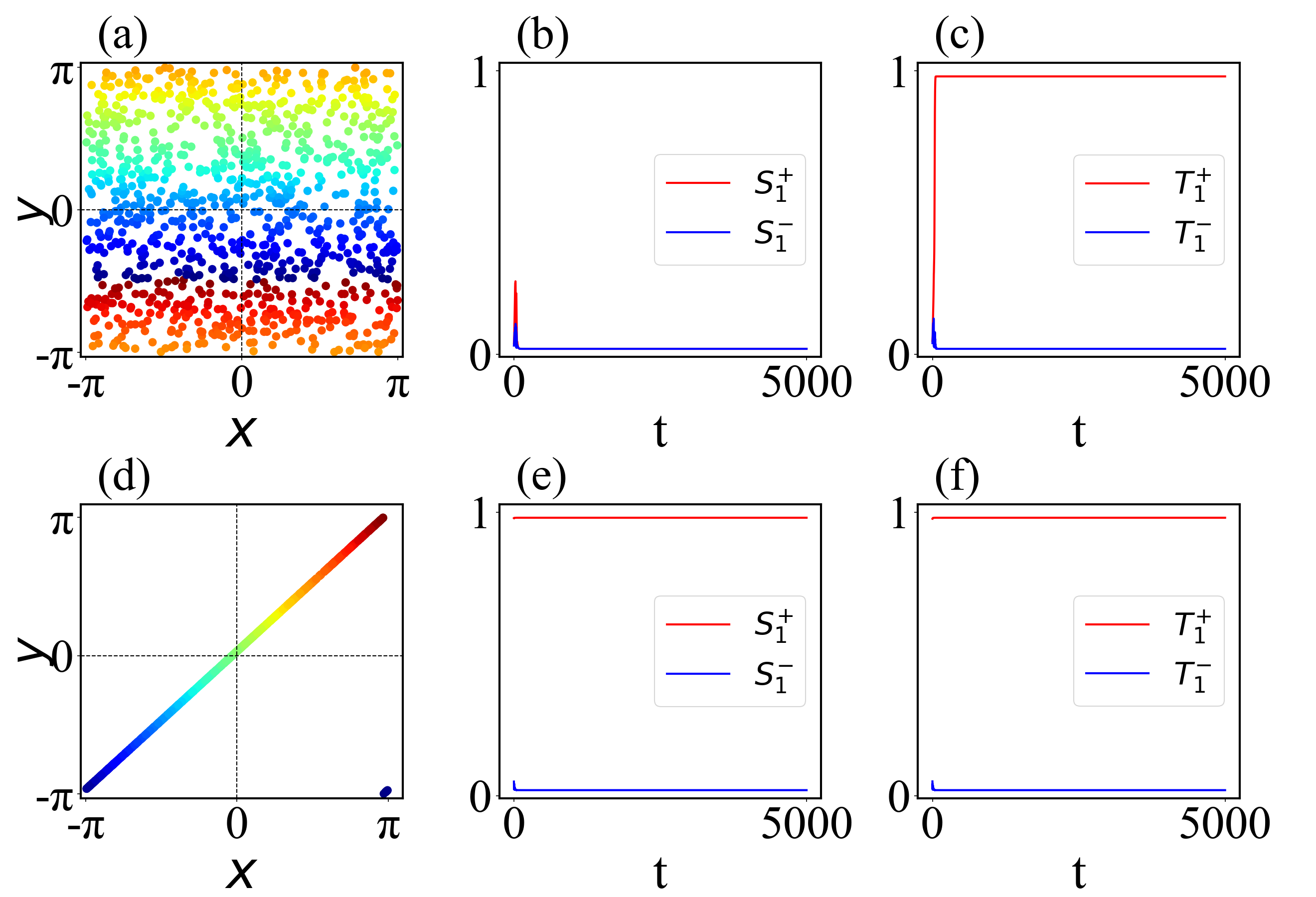}}
    \caption{Bistability between thick and thin phase wave state for $(J,K_1,K_2)=(1.5,-1,0.3)$. The Upper and lower rows correspond to the snapshots and time evolution of the order parameters for the thick and thin phase wave states, respectively.}
    \label{thick_thin_bistable}
\end{figure}

 Figure \ref{thick_thin_bistable} illustrates the bistability between the thick and thin phase wave states for $(J,K_1,K_2)=(1.5,-1,0.3)$. Here, the swarmalators self-organize to form the thick phase wave state when the initials are drawn randomly from the cube of length $2\pi$ (upper panel), while for initials chosen from the close neighborhood of the fixed point of the thin phase wave, the system exhibits thin phase wave state. 

\par All these bistability behaviors between the collective states of the system emerge as a direct consequence of the higher-order interactions. Such behavior was absent in the purely pairwise interaction systems, where multistability was not observed. Therefore, the inclusion of higher-order interactions plays a crucial role in inducing multistability in the swarmalator system, allowing different stable states to coexist, depending on the initial conditions. This highlights the significant impact of higher-order interactions in shaping the complex dynamical landscape of the swarmalator system.

\begin{figure*}[htp!] 
	\centerline
	{\includegraphics[width=\linewidth]{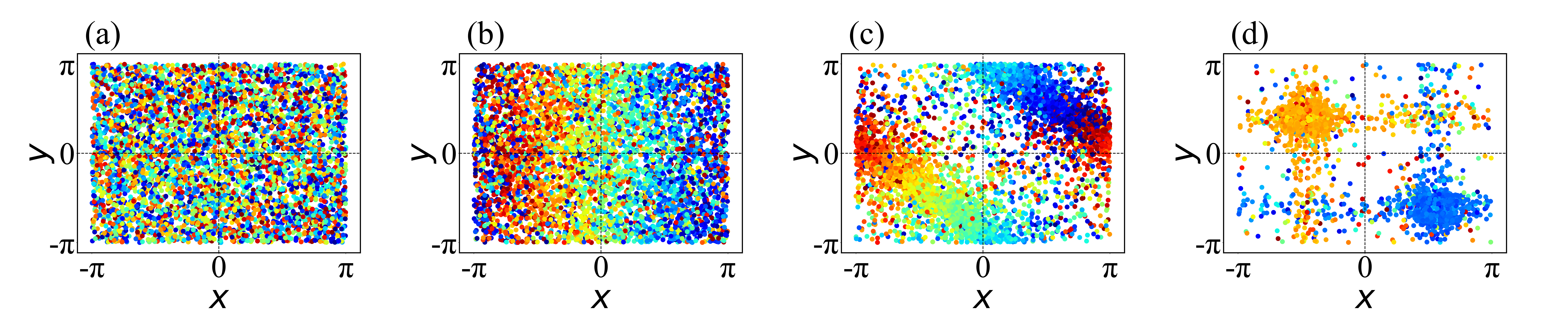}}
	\centerline
	{\includegraphics[width=\linewidth]{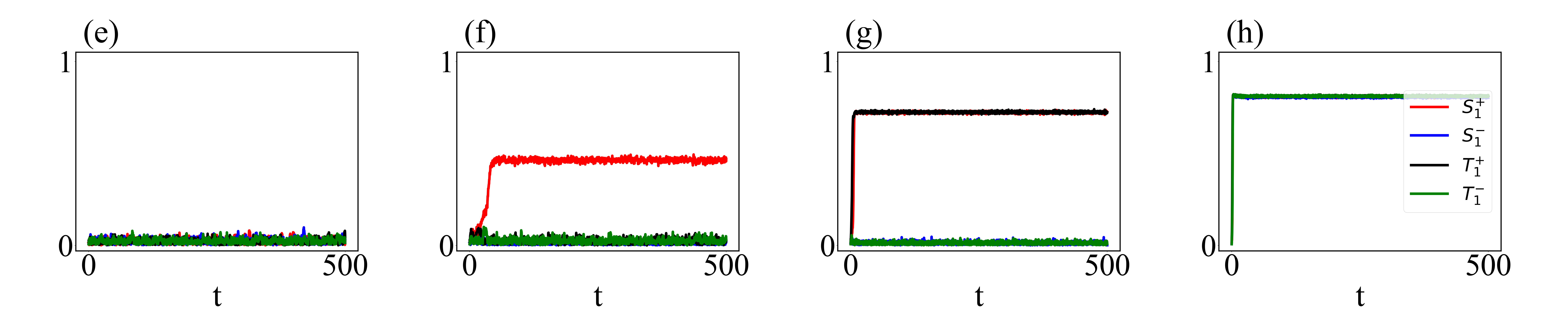}}
	\caption{\textbf{Collective states for nonidentical swarmalators.} Top row: snapshots of the states in $(x,y)$ plane with colors indicating the phase $\theta$, middle row: order parameters $S_{1}^{\pm}$ and $T_{1}^{\pm}$ as a function of time. (a, e) Async: $(J, K_1, K_2)=(15, -9, 10)$, (b, f) Thick phase wave: $(J, K_1, K_2)=(15, -7, 10)$, (c, g) Thin phase wave: $(J, K_1, K_2)=(15, -4, 10)$,  and (d, h) Sync: $(J, K_1, K_2)=(15, 2, 10)$. States are obtained by integrating Eq.~\eqref{model} with $N=5000$ swarmalators for a total of $T=500$ time units using an adaptive Julia Ode solver having relative tolerance of $10^{-8}$. The free velocities and natural frequencies of the swarmalators are drawn from Lorentzian distribution with mean zero and half-width $\Delta=1$.}
	\label{nonidentical_states}
\end{figure*}
\begin{figure}[hpt] 
	\centerline
	{\includegraphics[scale=0.15]{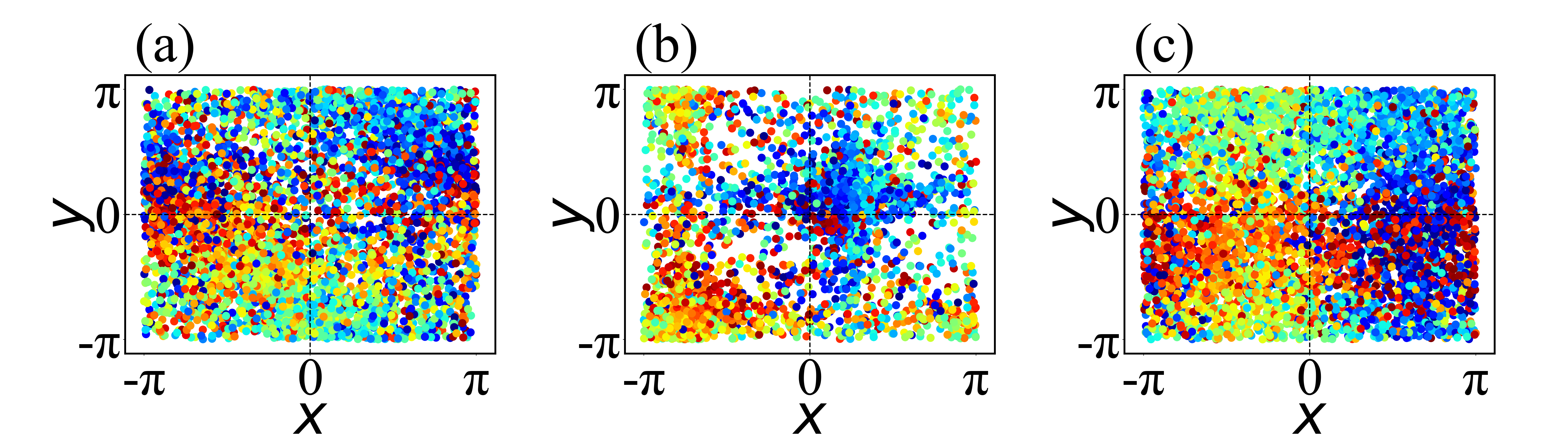}}
	\centerline
	{\includegraphics[scale=0.15]{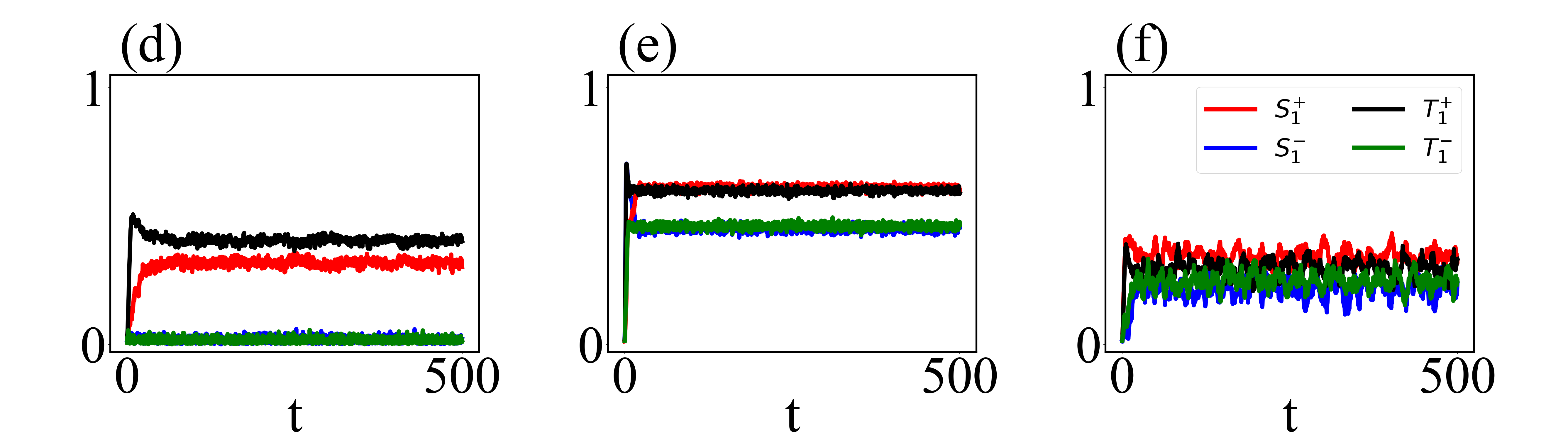}}
	\caption{\textbf{Mixed states} Top row: snapshots of the states in $(x,y)$ plane with colors indicating the phase $\theta$, middle row: order parameters $S_{1}^{\pm}$ and $T_{1}^{\pm}$ as a function of time. (a, d) Mixed Type I: $(J, K_1, K_2)=(15, -4.5, 1)$, (b, e) Mixed Type II: $(J, K_1, K_2)=(15, 1.35, 0.05)$, and (c, g) Mixed Type III: $(J, K_1, K_2)=(1.5, 10, 0.05)$. Other parameters are $(N,T,\Delta)=(5000,500,1)$.}
	\label{nonidentical_mixed_states}
\end{figure}

\section{Collective states for nonidentical swarmalators}\label{ni_states}
Here, we describe the collective states that arise from interactions among a population of nonidentical swarmalators. Similar to the case of identical swarmalators, we observe the emergence of async, sync, thin phase wave, and thick phase wave states (see Fig.~\ref{nonidentical_states}). Unlike the case of an identical population, the nonidentical swarmalators do not settle into a well-defined fixed-point configuration. Instead, they split into two subgroups: locked and drifting. For example, in the sync state, the slower swarmalators become locked in clusters, while the faster ones drift around these clusters. In the phase wave states, the division between the locked and drifting subgroups becomes even more intricate. As a result, the order parameter values achieve a maximal value of $0<S<1$ in the sync state, i.e., $(S_{1}^{+},S_{1}^{-},T_{1}^{+},T_{1}^{-})=(S,S,S,S)$. For the thin phase wave state, the order parameters take the form $(S_{1}^{+},S_{1}^{-},T_{1}^{+},T_{1}^{-})=(S,0,S,0)$, they are given by $(S_{1}^{+},S_{1}^{-},T_{1}^{+},T_{1}^{-})=(S,0,0,0)\; \text{or} \; (0,0,S,0)$.  

\par Figure \ref{nonidentical_mixed_states} portrays that the nonidentical swarmalators form a group of mixed states, which are basically a mixture of phase waves and sync states. These mixed states acts like transitional link between the phase waves states and sync state. Although their structure is difficult to depict clearly, the underlying process involves the symmetric phase wave developing an asymmetric deformation. Based on their role in the bifurcation sequence, these mixed states are categorized into three distinct types. Type I occurs as an intermediate state between the thick and thin phase wave states. Type II emerges at the junction of the thin phase wave and sync states. Type III is found between the thick phase wave and sync states. While these mixed states appear as intermediate states for small higher-order coupling, they do not emerge when higher-order interactions are sufficiently strong. In such cases, the system transitions directly between the phase wave states and the sync state, bypassing the intermediate mixed states. The characteristics of these states are more clearly understood through the values of the order parameters, as shown in the lower panel of Fig.~\ref{nonidentical_mixed_states}. In the Type I state, the order parameters are given by $(S_{1}^{+},S_{1}^{-},T_{1}^{+},T_{1}^{-})=(S_{1},0,S_{2},0)$. For the Type II state, the order parameters take the form $(S_{1}^{+},S_{1}^{-},T_{1}^{+},T_{1}^{-})=(S_{1},S_{2},S_{1},S_{2})$ and in the type III state, the order parameters are given by $(S_{1}^{+},S_{1}^{-},T_{1}^{+},T_{1}^{-})=(S_{1},S_{2},S_{3},S_{4})$, where $S_{i}>0$ $(i=1,2,3,4)$ are distinct.

\section{OA analysis for nonidentical swarmalators} \label{appendixb}
In the thermodynamic limit $N \rightarrow \infty$, the state of the system is defined by a probability density function $\rho(u,v,w,x_{\pm},y_{\pm},t)$ that denotes the probability to find a swarmalator at time $t$ having velocities $(u,v)$, intrinsic frequency $w$, and coordinates $x_{\pm}, y_{\pm}$. Then, $\rho$ satisfies the continuity equation 
\begin{equation}
	\dfrac{\partial \rho(u,v,w,x_{\pm},y_{\pm},t)}{\partial t} + \nabla \cdot (v \rho) = 0. \label{cont1}
\end{equation}
The OA ansatz here is a product of Poisson kernels
\begin{align}
	&\rho(u,v,w,x_{\pm}, y_{\pm},t) = \frac{1}{16 \pi^4} g(u) g(v) g(w) \nonumber \\
	& \times \Big[ 1 + \sum_{n=0}^{\infty} \alpha_x^n e^{i n x_+} + \text{c.c.} \Big] \times [1 + \sum_{m=0}^{\infty} \beta_x^m e^{i m x_-} + \text{c.c.} ] \nonumber \\
	& \times \Big[ 1 + \sum_{l=0}^{\infty} \alpha_y^l e^{i l y_+} + \text{c.c.} \Big]  \times [1 + \sum_{p=0}^{\infty} \beta_y^p e^{i p y_-} + \text{c.c.} ],
\end{align}
where $\alpha_{x,y}$ and $\beta_{x,y}$ are unknown quantities and are to be solved self-consistently. We substitute this ansatz into the continuity equation~\eqref{cont1} and then projection onto $e^{i x_{\pm}}, e^{i y_{\pm}}$ yields coupled complex differential equations of the Fourier modes for all harmonics $n,m,l$, and $p$,
	\begin{subequations}
		\begin{align}
			\dot{\alpha_x} = & -i u_+ \alpha_x +\frac{1}{2}\left[H_+^*(W_1^+,W_2^+)-H_+(W_1^+,W_2^+) \alpha_x^2\right]+\frac{\alpha_x}{2}\left[H_-^*(W_1^-,W_2^-) \beta_x^* - H_-(W_1^-,W_2^-) \beta_x\right] \nonumber \\ &+\frac{\alpha_x}{2}\left[G^*(Z_1^+,Z_2^+) \alpha_y^* - G(Z_1^+,Z_2^+) \alpha_y + G(Z_1^-,Z_2^-) \beta_y - G^*(Z_1^-,Z_2^-) \beta_y^*\right], 
		\end{align}
		\begin{align}
			\dot{\beta_x} = & -i u_- \beta_x+\frac{1}{2}\left[H_+^*(W_1^-,W_2^-)-H_+(W_1^-,W_2^-) \beta_x^2\right]+\frac{\beta_x}{2}\left[H_-^*(W_1^+,W_2^+) \alpha_x^* - H_-(W_1^+,W_2^+) \alpha_x\right] \nonumber \\ &+\frac{\beta_x}{2}\left[-G^*(Z_1^+,Z_2^+) \alpha_y^* + G(Z_1^+,Z_2^+) \alpha_y - G(Z_1^-,Z_2^-) \beta_y + G^*(Z_1^-,Z_2^-) \beta_y^*\right],
		\end{align}
		\begin{align}
			\dot{\alpha_y} = & -i u_+ \alpha_y +\frac{1}{2}\left[H_+^*(Z_1^+,Z_2^+)-H_+(Z_1^+,Z_2^+) \alpha_y^2\right]+\frac{\alpha_y}{2}\left[H_-^*(Z_1^-,Z_2^-) \beta_y^* - H_-(Z_1^-,Z_2^-) \beta_y\right] \nonumber \\ &+\frac{\alpha_y}{2}\left[G^*(W_1^+,W_2^+) \alpha_x^* - G(W_1^+,W_2^+) \alpha_x + G(W_1^-,W_2^-) \beta_x - G^*(W_1^-,W_2^-) \beta_x^*\right], 
		\end{align}
		\begin{align}
			\dot{\beta_y} = & -i u_- \beta_y+\frac{1}{2}\left[H_+^*(Z_1^-,Z_2^-)-H_+(Z_1^-,Z_2^-) \beta_y^2\right]+\frac{\beta_y}{2}\left[H_-^*(Z_1^+,Z_2^+) \alpha_y^* - H_-(Z_1^+,Z_2^+) \alpha_y\right] \nonumber \\ &+\frac{\beta_y}{2}\left[-G^*(W_1^+,W_2^+) \alpha_x^* + G(W_1^+,W_2^+) \alpha_x - G(W_1^-,W_2^-) \beta_x + G^*(W_1^-,W_2^-) \beta_x^*\right].
		\end{align}
		\label{alpha-beta}
	\end{subequations}
The expressions for the rainbow order parameters become
\begin{subequations}
	\begin{align}
		W_1^+ = S_1^+ e^{i \phi_1^+} = \int \alpha_{x}^*(u,v,w) g(u)g(v)g(w) du dv dw ,
	\end{align}
	\begin{align}
		W_1^- = S_1^- e^{i \phi_1^-} = \int \beta_{x}^*(u,v,w) g(u)g(v)g(w) du dv dw ,
	\end{align}
	\begin{align} 
		Z_1^+ = T_1^+ e^{i \psi_1^+} = \int \alpha_{y}^*(u,v,w) g(u)g(v)g(w) du dv dw ,
	\end{align}
	\begin{align} 
		Z_1^- = T_1^- e^{i \psi_1^-} = \int \beta_{y}^*(u,v,w) g(u)g(v)g(w) du dv dw .
	\end{align}
	\label{op1}
\end{subequations}

Important to note, these only hold on the sub-manifold $||\alpha_x|| = ||\beta_x|| = ||\alpha_y|| = ||\beta_y|| = 1$. The higher-order mixed harmonics $e^{i( n x_+ + m x_- + l y_+ + p y_-)}$ only close on this sub-manifold. The same structure has been reported earlier in the 1D swarmalator model \cite{yoon2022sync,anwar2024collective} and also in the periodic 2D swarmalator model with pairwise coupling~\cite{o2024solvable}. We use Eq.~\eqref{alpha-beta} and Eq.~\eqref{op1} to determine the expressions of the order parameters in the different emerging states as well as find their boundaries using them.

\section*{Acknowledgements}
M.S.A. and G.K.S. would like to thank the Department of Mathematics and the Namur Institute of Complex Systems, University of Namur, for support and hospitality during their research visit, where work on this article was carried out.


\end{document}

%% file: ex_shared.tex
\usepackage{lipsum}
\usepackage{amsfonts}
\usepackage{graphicx}
\usepackage{epstopdf}
\usepackage{algorithmic}
\ifpdf
  \DeclareGraphicsExtensions{.eps,.pdf,.png,.jpg}
\else
  \DeclareGraphicsExtensions{.eps}
\fi

\usepackage{enumitem}
\setlist[enumerate]{leftmargin=.5in}
\setlist[itemize]{leftmargin=.5in}

\newsiamremark{remark}{Remark}
\newsiamremark{hypothesis}{Hypothesis}
\crefname{hypothesis}{Hypothesis}{Hypotheses}
\newsiamthm{claim}{Claim}

\headers{A two-dimensional swarmalator model with higher-order interactions}{Md Sayeed Anwar, Gourab Kumar Sar, Timoteo Carletti and Dibakar Ghosh}

\title{A two-dimensional swarmalator model with higher-order interactions
\thanks{Submitted to the editors DATE.
}}

\author{
	Md Sayeed Anwar\thanks{Physics and Applied Mathematics Unit,
		Indian Statistical Institute, 203 B. T. Road, Kolkata 700108, India 
		(\email{sayeedanwar447@gmail.com}). These authors contributed equally to this work.}
	\and
	Gourab Kumar Sar\thanks{Physics and Applied Mathematics Unit, Indian Statistical Institute, 203 B. T. Road, Kolkata 700108, India (\email{mr.gksar@gmail.com}). These authors contributed equally to this work.}
	\and
	Timoteo Carletti\thanks{Department of Mathematics and Namur Institute for Complex Systems, naXys, University of Namur, 2 rue Grafé, Namur B5000, Belgium (\email{timoteo.carletti@unamur.be}).}
	\and
	Dibakar Ghosh\thanks{Physics and Applied Mathematics Unit, Indian Statistical Institute, 203 B. T. Road, Kolkata 700108, India (\email{diba.ghosh@gmail.com}).}
}

\usepackage{amsopn}
\usepackage{amssymb}
\usepackage{hyperref}
